\documentstyle[prb,aps,twocolumn]{revtex}
\input epsf
\begin{document}
\twocolumn[\hsize\textwidth\columnwidth\hsize
\csname@twocolumnfalse\endcsname
\draft
\title{Bose-Einstein statistics in thermalization 
and photoluminescence of quantum well excitons}
\author{A.L. Ivanov and P.B. Littlewood}
\address{University of Cambridge, Department of Physics, 
Theory of Condensed Matter Group, Cavendish Laboratory, 
Madingley Road, Cambridge CB3 0HE, UK}
\author{H. Haug}
\address{ Institut f\"ur Theoretische Physik, J.W. Goethe
Universit\"at Frankfurt, Robert-Mayer-Str. 8, D-60054 Frankfurt,
Germany }
\date{August 27, 1998}
\maketitle


\begin{abstract}
Quasi-equilibrium relaxational thermodynamics is developed to 
understand LA-phonon-assisted thermalization of Bose-Einstein 
distributed excitons in quantum wells. We study the quantum-statistical 
effects in the relaxational dynamics of the effective temperature of 
excitons $T = T(t)$. When $T$ is less than the degeneracy temperature 
$T_0$, well-developed Bose-Einstein statistics of quantum well excitons 
leads to nonexponential and density-dependent thermalization. 
At low bath temperatures $T_b \rightarrow 0$ the thermalization of 
quantum-statistically degenerate excitons effectively slows down and 
$T(t) \propto 1 / \ln t$. We also analyze the optical decay of 
Bose-Einstein distributed excitons in perfect quantum wells and show 
how nonclassical statistics influences the effective lifetime 
$\tau_{opt}$. In particular, $\tau_{opt}$ of a strongly degenerate gas 
of excitons is given by $2 \tau_R$, where $\tau_R$ is the intrinsic 
radiative lifetime of quasi-two-dimensional excitons. Kinetics of 
resonant photoluminescence of quantum well excitons during their 
thermalization is studied within the thermodynamic approach and taking 
into account Bose-Einstein statistics. We find density-dependent 
photoluminescence dynamics of statistically degenerate excitons. 
Numerical modeling of the thermalization and photoluminescence kinetics 
of quasi-two-dimensional excitons are given for GaAs/AlGaAs quantum wells. 
\end{abstract}
]
\narrowtext
\section{Introduction}
The fundamental features of relaxation and photoluminescence (PL) of 
the excitons in quantum wells (QWs) originate from the 
quasi-two-dimensionality (quasi-2D) of the system. In the last decade 
these basic processes in GaAs QWs have attracted continual attention 
\cite{Masumoto,Koteles,Feldmann,Damen,Deveaud1,Vinattieri,Kuhl,Wood,Weimann,Shah0,Shah1}. 
The formation of QW excitons through LO-phonon cascade emission, 
LA-phonon and carrier-carrier scattering completes within $20 \ ps$ 
after the initial excitation of electron-hole pairs 
\cite{Damen,Kuhl,Axt}. The created hot QW excitons then thermalize 
through low-energy acoustic phonon scattering. This process occurs 
in a sub-$ns$ time scale. In GaAs QWs the relaxation of hot excitons 
can be observed through resonant PL from the optically-active bulk 
modes which refer to a small in-plane momentum. Thus the rise and decay 
times of excitonic PL relate to the thermalization process. Furthermore, 
the PL kinetics of long-lifetime indirect excitons in a high-quality 
GaAs/AlGaAs coupled quantum well (CQW) is now the subject of 
experimental study \cite{Butov1}. There has been recent progress in the 
investigation of thermalization of quasi-2D excitons in ZnSe QWs 
\cite{Kalt}. In the latter case the picosecond LA-phonon-assisted 
kinetics of QW excitons is visualized through LO-phonon-assisted PL for 
all in-plane modes ${\bf p_{\|}}$. 

While the importance of interface disorder and the localization effects 
were recognized in the very first experimental \cite{Masumoto} and 
theoretical \cite{Takagahara} studies on the relaxation kinetics of QW 
excitons, the quality of GaAs QWs continuously improves from an 
inhomogeneous excitonic linewidth of about $7.5 \ meV$ \cite{Masumoto} 
toward a homogeneous one between $80$ and $120 \ \mu eV$ \cite{Wood}. 
In the present work we investigate LA-phonon-assisted relaxation 
kinetics in a perfect QW emphasizing the importance of Bose-Einstein 
(BE) statistics of quasi-2D excitons. In particular, we attribute the 
density-dependent PL kinetics reported, e.g., in 
Refs.~\cite{Damen,Kuhl}, to nonclassical statistics of QW excitons. 
In the past, most theoretical modeling of the relaxation kinetics in 
QWs \cite{Koteles,Feldmann,Wood,Shah0,Takagahara,Andreani1} dealt with 
a classical gas of Maxwell-Boltzmann distributed excitons. 
Quantum-statistical effects were included in numerical simulations of 
the relaxation kinetics of a trapped quasi-2D exciton gas \cite{Per1} 
and in the study of exciton-biexciton law of mass action in QWs 
\cite{Wolfe}. 

Crossover from classical to quantum statistics occurs near 
the degeneracy temperature $T_0$, given by 
\begin{equation} 
k_B T_0 = { 2 \pi \over g } \Bigg( { \hbar^2 \over M_x } \Bigg) 
\rho_{2D}  \ , 
\eqnum{1}
\label{intr1} 
\end{equation}
where $M_x$ is the in-plane translational mass of a QW exciton, 
$\rho_{2D}$ is the 2D concentration of excitons, and $g$ is the spin 
degeneracy factor of QW excitons. We put $g=4$, because for GaAs QWs 
the exchange interaction is rather weak \cite{Andreani2}. Furthermore, 
this interaction is completely suppressed for indirect excitons in 
CQWs. One has a classical gas of QW excitons at temperatures $T \gg 
T_0$. BE statistics smoothly develops with decreasing $T \sim T_0$ 
leading to occupation numbers $N_{\bf p_{\|}} \geq 1$ of the 
low-energy in-plane modes ${\bf p_{\|}}$. We will show that at helium 
temperatures nonclassical statistics affects the thermalization process 
already at moderate densities of QW excitons $\rho_{2D} 
\geq 3 - 5 \times 10^{10} \ cm^{-2}$. 

In this paper we develop {\it relaxational thermodynamics} and study 
within this approach how BE statistics influences thermalization and 
photoluminescence of quasi-2D excitons. Relaxational thermodynamics 
requires that exciton-exciton interaction is much stronger than 
exciton-LA-phonon coupling and is appropriate if the concentration 
$\rho_{2D}$ of QW excitons is larger than some critical density 
$\rho^c_{2D}$. In this case, the the exciton system establishes a 
quasi-equilibrium temperature $T$. For GaAs QWs we will get an estimate 
$\rho^c_{2D} \simeq 1 - 3 \times 10^9 \ cm^{-2}$.
Equation (\ref{term6}), which is the basic equation of the 
relaxational thermodynamics of QW excitons, provides us with an 
unified description of the thermalization process. The thermodynamic 
approach yields temporal evolution of the effective temperature of QW 
excitons $T = T(t)$ from the initial value $T_i = T(t=0)$ to the bath 
temperature $T_b$. While we study the thermalization process in the 
both limits, classical and degenerate, of a gas of QW excitons, the 
most interesting results refer to $T \leq T_0$. In this case one finds 
a density-dependent nonexponential relaxation of quasi-2D excitons. 
Both acceleration and slowing down of the thermalization kinetics may 
occur due to BE statistics. However, with the decreasing bath 
temperature $T_b$ slowing down of the relaxation starts to dominate in 
a quantum gas. In particular, for $T_b \rightarrow 0$ we derive 
$1/\ln t$ cooling law for QW excitons. 

The thermalization kinetics of excitons in GaAs QWs can be observed 
through resonant PL from the in-plane radiative modes. Therefore we 
generalize the PL theory \cite{Feldmann,Andreani1} to well-developed BE 
statistics of QW excitons and model numerically the PL process in GaAs 
QWs. It will be shown that at low temperatures nonclassical statistics 
changes the law $\tau_{opt} \propto T$, where $\tau_{opt}$ is the 
effective radiative lifetime of QW excitons. Furthermore, we calculate 
the temperature-independent corrections to the classical linear behavior 
of $\tau_{opt}(T)$. The corrections originate from BE statistics and can 
be traced even at high temperatures. 

Relaxational thermodynamics refers to the phonon-assisted 
thermalization kinetics of QW excitons. We will analyze the relaxation 
kinetics due to bulk LA-phonons, assuming the initial distribution of 
hot QW excitons is below the threshold for optical phonon emission. The 
bulk LA-phonons are due to a semiconductor substrate of a QW. 
While the lifetime of long wavelength acoustic phonons are in a 
sub-$\mu s$ time scale, the scattering LA-phonons leave and enter 
the QW area within a time $\sim L_z/v_s \sim 1 - 10 \ ps$ ($L_z$ 
is the thickness of a QW, $v_s$ is the longitudinal sound velocity). 
Therefore, we assume the Planck distribution of the LA-phonons 
interacting with QW excitons. The LA-phonon-assisted kinetics will 
be treated for an isolated band of ground-state QW excitons. 
Being applied to heavy-hole excitons in GaAs QWs with $L_z \leq 
100 \ \AA$ this assumption means QW exciton energies $E \leq 
10 \ meV$ and temperatures less than $100 \ K$. 
Only the exciton-LA-phonon deformation potential interaction 
is included in our model. 

The in-plane momentum is conserved in the scattering of QW excitons 
by bulk LA-phonons. The momentum conservation in the $z$-direction 
(the QW growth direction) is relaxed. As a result, a scattered QW 
exciton interacts with a continuum spectral band of scattering 
LA-phonons of a given direction of propagation. In contrast, an exciton 
in bulk semiconductors couples in Stokes or anti-Stokes scattering 
only with one phonon mode of a given direction. In Fig.~1 we depict 
a schematic picture of the states which are involved in the scattering 
of a QW exciton with in-plane momentum ${\bf p_{\|}} = 0$. The energy 
state $E=0$ couples to the continuum energy states $E \geq E_0 = 2 M_x 
v_s^2$, i.e., to the QW states which lie inside the acoustic cone given 
by $E = E(p_{\|}) = \hbar v_s p_{\|}$. The energy $E_0$ is an important 
parameter of the relaxational thermodynamics of QW excitons. For GaAs 
QWs with $M_x = 0.3 \ m_0$ ($m_0$ is the free electron mass) and $v_s = 
3.7 \times 10^5 \ cm/s$ one has $E_0 = 46.7 \ \mu eV$ and $E_0/k_B = 
0.54 \ K$. 
\begin{figure}[t]
\begin{center}
\leavevmode
\epsfxsize=8cm \epsfbox{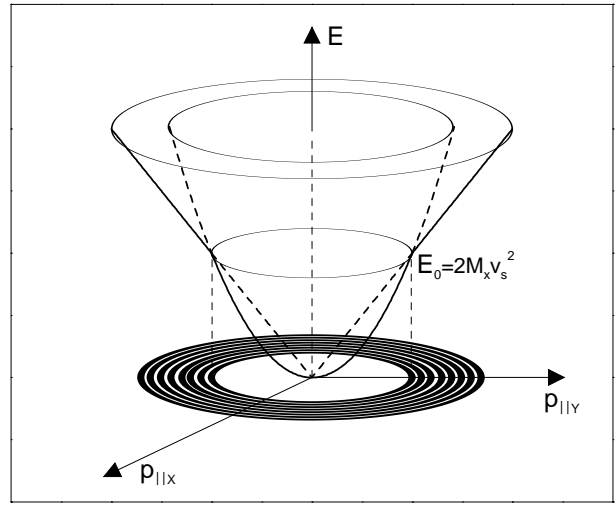}
\end{center}
\caption{Schematic picture of the energy-momentum conservation in 
scattering of QW excitons by bulk LA-phonons. The parabolic and conical 
surfaces refer to the excitonic $E = \hbar^2 p_{\|}^2 /2M_x$ and 
acoustic $E = \hbar v_s p_{\|}$ dispersions, respectively. The bold 
rings show the states $\hbar p_{\|} \geq 2 M_x v_s$ of QW excitons 
which couple to the ground-state mode ${\bf p}_{\|} = 0$.}
\label{fig1b}
\end{figure}

The relaxation kinetics of QW excitons coupled to thermal bulk 
LA-phonons is given by the Boltzmann equation (see, e.g., \cite{Per2}): 
\begin{eqnarray}
{\partial \over \partial t} N_{\varepsilon'}
& = & - {2 \over \tau_{sc}} 
\int_0^{\infty} d\varepsilon \ \varepsilon^2 \int_{-1}^{1} 
du \ |F_z(a \varepsilon u)|^2 
\nonumber \\
& \times &
\Bigg\{ 
{1 \over \sqrt{4\varepsilon' (1-u^2) - (\varepsilon + 1 - 
\varepsilon u^2)^2} } 
\nonumber \\
& \times & \Big[ N_{\varepsilon'} (1+n^{ph}_{\varepsilon}) 
(1+N_{\varepsilon'-\varepsilon})  
-(1+N_{\varepsilon'}) n^{ph}_{\varepsilon}
N_{\varepsilon'-\varepsilon} \Big]  
\nonumber \\
& \times & 
\Theta \Big( 2\sqrt{\varepsilon'(1-u^2)} - \varepsilon - 1 + 
\varepsilon u^2 \Big) 
\Theta(\varepsilon'-\varepsilon) \nonumber \\ 
& + & \ {1 \over \sqrt{4\varepsilon' (1-u^2) - 
(\varepsilon - 1 -  \varepsilon u^2 )^2}} 
\nonumber \\
& \times & \Big[ N_{\varepsilon'} n^{ph}_{\varepsilon} 
(1+N_{\varepsilon'+\varepsilon}) - (1+N_{\varepsilon'}) 
(1+n^{ph}_{\varepsilon}) N_{\varepsilon'+\varepsilon} \Big] 
\nonumber \\
& \times & 
\Theta \Big( 2\sqrt{\varepsilon' (1-u^2)} + \varepsilon - 1 -  
\varepsilon u^2 \Big) 
\nonumber \\
& \times &
\Theta \Big( 2\sqrt{\varepsilon' (1-u^2)} - \varepsilon + 1 + 
\varepsilon u^2 \Big) \Bigg\} \ ,
\eqnum{2}
\label{intr2}
\end{eqnarray}
where the dimensionless energy is $\varepsilon = E/E_0 = 
E/(2M_x v_s^2)$, $n^{ph}_{\varepsilon} = 1/[\exp(\varepsilon E_0/k_B 
T_b) - 1]$ and $N_{\varepsilon}$ are the distribution functions of bulk 
LA-phonons and QW excitons, respectively, $\Theta$ is the Heaviside 
step-function. The integration variable $u$ is given by $u = \cos 
\theta$, where $0 \leq \theta \leq \pi$ is the angle between $z$-axis 
and the wavevector of a scattering bulk LA-phonon. The scattering time is 
$\tau_{sc} = (\pi^2 \hbar^4 \rho)/(D^2 M_x^3 v_s)$, where $\rho$ is the 
crystal density and $D$ is the deformation potential. The form-factor 
$F_z(\chi) = [\sin(\chi)/ \chi] [e^{i \chi} / (1 - \chi^2/\pi^2)]$ 
refers to an infinite QW confinement potential \cite{Bockelmann}. 
This function describes the relaxation of the momentum conservation 
in the $z$-direction and characterizes a spectral band of LA-phonons, 
which effectively interact with a QW exciton. The dimensionless 
parameter $a \sim 1$ is defined by $a = (L_z M_x v_s)/\hbar$. 

The kinetic Eq.~(\ref{intr2}) deals with an isotropic in-plane 
distribution of QW excitons, i.e., the occupation number $N_{\bf p_{\|}}$ 
of the in-plane mode ${\bf p_{\|}}$ relates to the distribution function 
$N_E$ by $N_{\bf p_{\|}} = N_{E_{\bf p_{\|}}} = N_E$. This approximation 
corresponds to the experimental conditions of Refs.~[1-10]. The first 
(second) term in the figure brackets on the right-hand side (r.h.s.) of 
Eq.~(\ref{intr2}) describes the phonon-assisted Stokes (anti-Stokes) 
scattering. In accordance with Eq.~(\ref{intr2}), the relaxation kinetics 
into the ground-state mode ${\bf p_{\|}} = 0$ is given by 
\begin{eqnarray} 
{\partial \over \partial t} N_{E=0} & = & {2 \pi \over \tau_{sc}} \ 
\int_1^{\infty} \ d \varepsilon \ \varepsilon \ \sqrt{\varepsilon 
\over \varepsilon - 1} \ \ \Big|F_z \Big( a 
\sqrt{\varepsilon(\varepsilon - 1)} \Big) \Big|^2 
\nonumber \\
& \times & \ \left[ N_{\varepsilon} (1 + 
n^{ph}_{\varepsilon}) - N_{E=0} (n^{ph}_{\varepsilon} - 
N_{\varepsilon}) \right] \ .  
\eqnum{3}
\label{intr3}
\end{eqnarray}
The integral on the r.h.s. of Eq.~(\ref{intr3}) characterizes the 
coupling of the ground-state mode $E=0$ to continuum of the energy 
states $E \geq E_0$ (see Fig.~1). 

In Sec.~II, we develop the relaxational thermodynamics of QW excitons 
coupled to bulk LA-phonons. A basic equation for the effective 
temperature $T$ of quasi-equilibrium QW excitons is derived from 
LA-phonon-assisted kinetics given by Eq.~(\ref{intr2}). The conditions 
of the validity of the thermodynamic picture are analyzed and tested 
for excitons in GaAs QWs. 

In Sec.~III, the thermalization law $T = T(t)$ is studied for both 
classical ($T \gg T_0$) and well-developed BE statistics ($T \ll T_0$) 
of quasi-equilibrium QW excitons. We demonstrate that BE statistics 
strongly influences the thermalization process and leads to the 
density-dependent characteristic thermalization time $\tau_{th} = 
\tau_{th}(\rho_{2D})$ for $|T - T_b| \ll T_b$ and to the 
{\it nonexponential} density-dependent relaxation at $T - T_b \geq 
T_b$. In particular, for low bath temperatures $T_b \leq E_0/k_B$ we 
find a very slow thermalization $T(t) \propto 1/\ln t$ of quantum 
degenerate quasi-2D excitons. The numerical simulations of the 
relaxational dynamics are given for excitons in GaAs QWs. 

In Sec.~IV, we develop a theory of resonant PL of statistically 
degenerate QW excitons. An effective radiative lifetime $\tau_{opt}$ 
of a quasi-2D excitonic gas is calculated for BE-distributed QW 
excitons. We show that the law $\tau_{opt} \propto T$, which is valid 
for classical statistics of QW excitons, is violated at low temperatures 
and at $T \rightarrow 0$ the effective decay time $\tau_{opt}$ tends 
to $2 \tau_R$, where $\tau_R$ is the intrinsic radiative lifetime of 
a QW exciton. The PL kinetics of QW excitons is described in the 
thermodynamic approach within three coupled equations for $T(t)$, 
$\rho_{2D}(t)$, and $\tau_{opt}(T,\rho_{2D})$. We give numerical 
modeling of the $T$- and $\rho_{2D}$-dependent PL kinetics of quantum 
degenerate quasi-2D excitons. 

In Sec.~V, we discuss the influence of interface polaritons and QW 
biexcitons on the relaxation kinetics of statistically degenerate 
excitons in perfect QWs. Our theory is always appropriate for 
thermalization of indirect excitons in CQWs, where even at small 
densities $\rho_{2D}$ the interface polariton effect is rather weak 
and biexciton states are unbound. For single QWs we argue that QW 
polaritons and biexcitons are considerably weakened at large densities, 
due to particle-particle scattering. In this case our model is 
applicable to usual QWs with direct excitons. 

In Appendix A, in order to estimate the critical density $\rho_{2D}^c$ 
of QW excitons for the development of relaxational thermodynamics we 
calculate the characteristic equilibration time $\tau_{x-x}$ due to 
exciton-exciton scattering. The equilibration rate $1/\tau_{x-x}$ is 
found in both limits of classical and well-developed BE statistics. 

In Appendix B, we analyze how a relatively large homogeneous linewidth 
$\sim \hbar / \tau_{x-x} \gg \hbar / \tau_{th}, \ \hbar / \tau_R$ of 
high-density QW excitons in a perfect QW influences the PL process. 

\section{Relaxational Thermodynamics of QW Excitons}
In this section we summarize the thermodynamic relations for an 
ideal two-dimensional gas of bosons and derive the basic equation for 
relaxational dynamics of QW excitons. 
\subsection{Thermodynamic relations for quasi-2D excitons}
The thermodynamic equation $\mu = \mu(T,\rho_{2D})$ for a 
quasi-equilibrium gas of QW excitons can be derived from the 
condition: 
\begin{equation} 
\rho_{2D} = {1 \over S} \sum_{\bf p_{\|}} N_{\bf p_{\|}}^{eq} = 
{2 g M_x k_B T \over \pi \hbar^2} \int_0^{\infty} { d z \over 
e^{-\mu/k_B T} e^z - 1 } \ .
\eqnum{4}
\label{term1} 
\end{equation}
From Eq.~(\ref{term1}) one obtains 
\begin{equation} 
\mu = k_B T \ln \left( 1 - e^{-T_0/T} \right) \ ,
\eqnum{5}
\label{term2} 
\end{equation}
where the degeneracy temperature $T_0$ is given by Eq.~(\ref{intr1}) 
with $g=4$. With the chemical potential $\mu$ of Eq.~(\ref{term2}) the 
equilibrium distribution function of QW excitons is 
\begin{equation}
N^{eq}_E = {1 - e^{-T_0/T} \over e^{E/k_BT} + e^{-T_0/T} - 1} \ . 
\eqnum{6}
\label{term3} 
\end{equation}
In particular, the occupation number of the ground-state mode 
is given by $N^{eq}_{\bf p_{\|}=0} = N^{eq}_{E=0} = \exp(T_0/T) - 1$. 

Classical, Maxwell-Boltzmann, statistics of QW excitons is 
realized for $T \gg T_0$. In this case Eqs.~(\ref{term2})-(\ref{term3}) 
reduce to 
\begin{equation} 
\mu = k_B T \ln(T_0/T) \ \ \ \ \mbox{and} \ \ \ \ N^{eq}_{E=0} = 
T_0/T \ll 1 \ ,
\eqnum{7}
\label{term4} 
\end{equation}
i.e., the occupation numbers of QW modes ${\bf p_{\|}}$ are much less 
than unity. In the opposite limit $T < T_0$ of well-developed BE 
statistics one has 
\begin{equation} 
\mu = - k_B T e^{-T_0/T} \ \ \ \ \mbox{and} \ \ \ \ N^{eq}_{E=0} = 
e^{T_0/T} \gg 1 \ . 
\eqnum{8}
\label{term5} 
\end{equation}
According to Eq.~(\ref{term5}), the chemical potential of a degenerate 
gas of QW excitons approaches zero much faster than the temperature 
$T$. While the chemical potential $\mu < 0$ for $T > 0$ and the BE 
condensation of the QW excitons is absent within the thermodynamic 
approach for temperatures above zero, the occupation number of 
the ground-state mode ${\bf p_{\|}}$ increases exponentially with 
decreasing $T \ll T_0$. 

Equations (\ref{term2})-(\ref{term5}) can be applied to an 
arbitrary two-dimensional quasi-ideal gas of Bose-particles 
with quadratic dispersion. The specific characteristics of 
the bosons, like the spin degeneracy factor $g$ and the 
translational mass $M_x$, enter the thermodynamic relationships 
only through the degeneracy temperature $T_0$ defined by 
Eq.~(\ref{intr1}). 

For GaAs QWs one estimates $T_0$ $= (\pi \hbar^2 \rho_{2D})/(2M_xk_B)$ $= 
4.6 \ K$ ($k_B T_0 = 399.5 \ \mu eV$) for $\rho_{2D} = 10^{11} \ 
cm^{-2}$. This density of quasi-2D excitons corresponds to the mean 
interparticle distance $\sim 1/\sqrt{\rho_{2D}} \simeq 320 \ \AA$ and 
to the Mott parameter $\rho_{2D} [a_x^{(2D)}]^2 \simeq 0.04$, i.e., 
the QW excitons are still well-defined quasiparticles. Here 
$a_x^{(2D)}$ is the Bohr radius of a quasi-2D exciton. In next 
subsection we will use Eqs.~(\ref{term2})-(\ref{term5}) 
to develop the relaxational dynamics. 
\subsection{Thermalization equation for quasi-equilibrium excitons
in QWs}
The thermalization of QW excitons occurs through a sequence of 
quasi-equilibrium thermodynamic states, which are characterized 
by an effective temperature $T = T(t)$ and chemical potential 
$\mu = \mu(t)$, provided that the exciton-exciton interaction in 
a QW is much stronger than the coupling of QW excitons with bulk 
LA-phonons. The initial interaction is conservative and equilibrates 
the system of QW excitons without change of total energy. 
The characteristic equilibration time $\tau_{x-x}$ depends on 
the density $\rho_{2D}$ of QW excitons. For excitons distributed below 
the threshold for LO-phonon emission, energy (effective temperature) 
relaxation results from QW exciton -- bulk LA-phonon scattering and 
is characterized by an effective thermalization time $\tau_{th}$. 
The hierarchy of interactions means that in a large and interesting 
range of density we have $\tau_{x-x} \ll \tau_{th}$. In this case the 
equilibration and thermalization kinetics can be naturally separated 
(see, e.g., Ref.~\cite{Levich}) and the thermodynamic approach is 
correct. 

The condition $\tau_{x-x} \ll \tau_{th}$ will hold for $\rho_{2D} 
> \rho^c_{2D}$, where $\rho^c_{2D}$ is some critical density for QW 
excitons. For a classical distribution of quasi-2D excitons one has 
$T_0/T \ll 1$ and $N^{eq}_E = (T_0/T) \exp(-E/k_BT) \ll 1$, according 
to Eq.~(\ref{term3}). In this limit the characteristic equilibration 
time $\tau_{x-x}$ is estimated by Eq.~(A4) of Appendix~A, while the 
thermalization time $\tau_{th}$ is given by Eq.~(\ref{exp4}) of 
section III. The comparison of Eq.~(A4) and Eq.~(\ref{exp4}) yields 
$\tau_{x-x} \ll \tau_{th}$ provided that 
\begin{equation} 
k_B T_0 > k_B T^c_0 = 
{4 \hbar \over \pi} \Bigg( {\mu_x \over M_x} \Bigg)^2 \ 
{ C_{2D} \over \tau_{sc} } \ , 
\eqnum{9}
\label{term7} 
\end{equation}
where $\mu_x$ is the reduced mass of a QW exciton and the constant 
$C_{2D} = C_{2D}(a) \gg 1$ is defined by Eq.~(\ref{exp4}). For a 
classical gas of excitons in GaAs QW of thickness $L_z = 100 \ \AA$ 
we estimate $\rho^c_{2D} = 1.2 \times 10^9 \ cm^{-2}$ and the 
corresponding temperature scale $T^c_0 = 56 \ mK$. For comparison, 
the critical density of about $4 \times 10^9 \ cm^{-2}$ has been 
estimated in experiments \cite{Damen} with high-quality GaAs QWs of 
$L_z = 80 \ \AA$. For well-developed BE statistics of QW excitons one 
has $T_0/T \gg 1$ and the occupation numbers of the energy states $E 
\geq k_B T \exp(-T_0/T)$ are given by the Planck function, i.e., 
$N^{eq}_{E>0} = 1 / [\exp(E/k_BT) - 1]$, according to 
Eq.~(\ref{term3}). In this case $\tau_{x-x}$ and $\tau_{th}$ are given 
by Eq.~(A5) of Appendix and Eq.~(\ref{exp7}) of section III, 
respectively. The condition $\tau_{x-x} \ll \tau_{th}$ reduces to 
\begin{equation} 
E_0 \Bigg( {T_0 \over T} \Bigg) e^{T_0/T} > 
{4 \hbar \over (1 - \pi/4)} \Bigg( {\mu_x \over M_x} \Bigg)^2 \ 
{ {\widetilde C}_{2D} \over \tau_{sc} } \ , 
\eqnum{10}
\label{term8} 
\end{equation}
where the constant ${\widetilde C}_{2D} = {\widetilde C}_{2D}(a) 
\gg 1$ is given by Eq.~(\ref{exp7}). The inequality (\ref{term8}) 
always holds for strongly degenerate QW excitons. 

The criteria (\ref{term7}) and (\ref{term8}) of the validity of the 
thermodynamic picture at $T \gg T_0$ and $T \ll T_0$, respectively, 
are independent of the scattering length $\sim a_x^{(2D)}$ of 
exciton-exciton interaction. This is due to the 
quasi-two-dimensionality of QW excitons (for details see Appendix~A). 
In contrast, in a three-dimensional gas the equilibration time 
$\tau_{x-x}$ due to particle-particle interaction explicitly depends 
on the scattering length \cite{Landau}. In further analysis we assume 
the hierarchy of interactions, i.e., that $\rho_{2D} > \rho^c_{2D}$ 
so that the inequality (\ref{term7}) is valid. The initial density of 
photogenerated excitons in GaAs QWs is usually larger than 
$10^9 \ cm^{-2}$ [3-10], indicating that the thermodynamic 
picture of relaxation is adequate for typical experimental conditions. 
We will omit the superscript in $N^{eq}_E$, because within our 
approach $N_E = N_E^{eq}$. 

The thermalization dynamics of QW excitons is determined by the slowest 
elementary LA-phonon-assisted relaxation process of the kinetic 
Eqs.~(\ref{intr2}) and (\ref{intr3}). The joint density of states for 
Stokes and anti-Stokes scattering from the energy mode $E$ continuously 
decreases with decreasing $E$, according to the r.h.s. of 
Eq.~(\ref{intr2}) (note, that the density of quasi-2D excitonic states 
is constant given by $16 \pi^2 M_x / \hbar^2$ for $g=4$). Moreover, 
the low-energy states $E \leq E_0/4$ are not active in Stokes 
scattering and couple with the corresponding LA-phonon-separated 
states $E \geq E_0$ only through the anti-Stokes process. The above 
arguments show a particular status of Eq.~(\ref{intr3}), which 
describes the relaxation kinetics into the ground-state mode. 
Furthermore, the ground-state mode ${\bf p_{\|}} = 0$ refers 
to the lowest energy $E=0$ and to the largest occupation number 
$N_{E=0}$. For well-developed BE statistics of QW excitons 
one has $N_{E=0} \gg 1$. As a result, population of the state 
$E=0$ generally requires an additional time in comparison with 
that for the high-energy states $E \gg k_B T$ with $N_E \ll 1$. 
Thus LA-phonon-assisted occupation of the ground-state mode 
is the slowest, ``bottleneck'', relaxation process, which 
determines the thermalization kinetics of quasi-equilibrium QW 
excitons. A similar picture holds for thermalization of excitons 
in bulk semiconductors \cite{Ivanov0,Ivanov2}. 

With the substitution of $N_E = N_E^{eq}$ given by Eq.~(\ref{term3}) 
in Eq.~(\ref{intr3}) we derive 
\begin{eqnarray}
{\partial \over \partial t} T & = & - { 2 \pi \over \tau_{sc} } 
\left( {T^2 \over T_0} \right) 
\Big(1 - e^{-T_0/T} 
\Big) \int_1^{\infty} d \varepsilon \ \varepsilon 
\sqrt{\varepsilon \over \varepsilon - 1} 
\nonumber \\
& \times & \Big|F_z \Big(a 
\sqrt{\varepsilon(\varepsilon - 1)} \Big)\Big|^2  
{ e^{\varepsilon E_0 / k_B T_b} - e^{\varepsilon E_0 / k_B T} 
\over (e^{\varepsilon E_0 / k_B T} 
+ e^{-T_0/T} - 1) } 
\nonumber \\
& \times &
{ 1 \over (e^{\varepsilon E_0 / k_B T_b} - 1) } \ .  
\eqnum{11}
\label{term6} 
\end{eqnarray}
Equation~(\ref{term6}) describes the thermalization dynamics $T = T(t)$ 
of QW excitons from the effective temperature $T_i = T(t=0)$ to the 
bath temperature $T_b$. A finite lifetime $\tau'$ of excitons, due to 
radiative and nonradiative recombination, can be incorporated in 
Eq.~(\ref{term6}) by the degeneracy temperature $T_0(t) \propto 
\rho_{2D}(t=0) \exp(-t/\tau')$. In next section we apply 
Eq.~(\ref{term6}) to the thermalization kinetics of QW excitons in the 
absence of their recombination ($\tau' \rightarrow \infty$). This 
analysis refers to the case when the characteristic thermalization time 
$\tau_{th} \ll \tau'$. In section IV we will use Eq.~(\ref{term6}) in 
order to develop a theory of resonant photoluminescence of quantum 
degenerate excitons in perfect QWs, when $\tau' = \tau_{opt}$ is 
determined by the intrinsic radiative lifetime $\tau_R$ of QW excitons. 
\section{Thermalization kinetics of quasi-equilibrium QW excitons}
In this section the thermalization dynamics of QW excitons is analyzed 
within Eq.~(\ref{term6}) for $\rho_{2D}=const$. First we linearize 
Eq.~(\ref{term6}) about some bath temperature $T_b$ to study the 
exponential relaxation of QW excitons with the effective temperature 
$|T - T_b| \ll T_b$. Linearization is not appropriate for the 
thermalization kinetics at $T_b = 0$. Here we show that the 
thermalization becomes very slow, with $T(t) \propto 1/\ln t$. In the 
last subsection, the cooling of hot QW excitons with $T - T_b \geq T_b$ 
is treated for both classical and strongly degenerate limits of BE 
statistics. There are four characteristic temperature scales, $T$, 
$T_0$, $T_b$, and $E_0/k_B$, and the precise form of the relaxation will 
depend on all four. Note however that if $T \gg T_0$, the degeneracy 
temperature is not a relevant parameter, and the most relevant aspects 
of phonon bottleneck effects are captured by the ratio $E_0/k_BT_b$. 
In practice the relaxational dynamics is controlled by two parameters, 
$T_0/T$ and $E_0/k_BT_b$. 
\subsection{Relaxation kinetics between nearby thermodynamic 
states}
If the effective temperature $T$ of QW excitons is close to the 
bath $T_b$, the basic thermodynamic Eq.~(\ref{term6}) reduces to 
\begin{equation} 
{\partial \over \partial t} T = - { 1 \over \tau_{th} } 
(T - T_b) \ ,  
\eqnum{12}
\label{exp1} 
\end{equation}
where the effective thermalization time $\tau_{th}$ is given by 
\begin{eqnarray}
{1 \over \tau_{th} } & = & \left( {2 \pi \over \tau_{sc}} \right) 
\left( {E_0 \over k_B T_0} \right) \Big(1 - e^{-T_0/T_b} 
\Big) \int_1^{\infty} d \varepsilon \ \varepsilon^2 
\sqrt{\varepsilon \over \varepsilon - 1} 
\nonumber \\
& \times & \Big|F_z \Big(a 
\sqrt{\varepsilon(\varepsilon - 1)} \Big)\Big|^2  
{ 1 \over (e^{\varepsilon E_0 / k_B T_b} 
+ e^{-T_0/T_b} - 1) } 
\nonumber \\
& \times & { e^{\varepsilon E_0 / k_B T_b} 
\over (e^{\varepsilon E_0 / k_B T_b} - 1) } \ . 
\eqnum{13}
\label{exp2} 
\end{eqnarray} 
The linear approximation of Eq.~(\ref{term6}) by Eq.~(\ref{exp1}) is 
valid for $|T - T_b|/T_b < k_BT_b/E_0$ and can be done for any 
$T_b > 0$. Equation~(\ref{exp1}) corresponds to the {\it exponential} 
thermalization law $T(t) = T_b + \Delta T \exp( -t/\tau_{th} )$, 
where $\Delta T = T_i - T_b$ and $T_i = T(t=0)$ is the initial 
temperature of QW excitons. The thermalization time $\tau_{th}$ 
uniquely describes all quasi-equilibrium relaxation processes in 
a gas of quasi-2D excitons. For example, the relaxation kinetics 
into the ground-state mode $E=0$ is given by $N_{E=0}(t) = N^f_{E=0} 
+ (N^i_{E=0} - N^f_{E=0}) \exp(-t/\tau_{th})$, where $N^i_{E=0} = 
\exp(T_0/T_i) - 1$ and $N^f_{E=0} = \exp(T_0/T_b) - 1$. We will 
analyze Eq.~(\ref{exp2}) in the limit of classical and 
well-developed BE statistics of QW excitons, respectively. 

1. {\it Classical gas of QW excitons ($T_b \gg T_0$). } 

In this case Eq.~(\ref{exp2}) yields 
\begin{eqnarray}
{1 \over \tau_{th} } & = & \left( {2 \pi \over \tau_{sc}} \right) 
\left( {E_0 \over k_B T_b} \right) \int_1^{\infty} d \varepsilon \ 
\varepsilon^2 \sqrt{\varepsilon \over \varepsilon - 1} 
\nonumber \\ 
& \times & 
{ \Big|F_z \Big(a \sqrt{\varepsilon(\varepsilon - 1)} \Big)\Big|^2  
\over \Big( e^{\varepsilon E_0 / k_B T_b} - 1 \Big) } \ . 
\eqnum{14}
\label{exp3} 
\end{eqnarray} 
From Eq.~(\ref{exp3}) one concludes, as expected, that the characteristic 
thermalization time of the Maxwell-Boltzmann distributed QW excitons 
is indeed independent of their concentration $\rho_{2D}$. In Fig.~2 
we plot $\tau_{th} = \tau_{th}(T_b)$ calculated by Eq.~(\ref{exp3}) 
for GaAs QW of thickness $L_z = 100 \ \AA$. 
The ratio $E_0 / k_B T_b$ determines the high- and 
low-temperature limits of $\tau_{th}$. 
\begin{figure}[t]
\begin{center}
\leavevmode
\epsfxsize=8cm \epsfbox{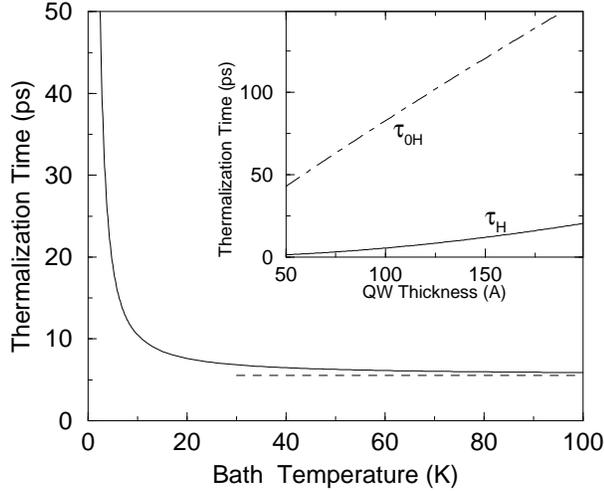}
\end{center}
\caption{ The thermalization time $\tau_{th} = \tau_{th}(T_b)$ of a 
classical gas of QW excitons. GaAs QW with $L_z$ $= 100 \ \AA$ and 
$\tau_{sc}$ $= 13.9 \ ns$ [see Eq.~(\ref{exp3})]. Inset: the 
high-temperature thermalization time $\tau_H = \tau_H(L_z) \propto 
L_z^2$ of Maxwell-Boltzmann distributed QW excitons (solid line) and 
the high-temperature thermalization time $\tau_{0H} = \tau_{0H}(L_z) 
\propto L_z$ of strongly statistically degenerate QW excitons (
dot-dashed line). The latter dependence refers to $T_0 = 
(k_B T_b^2)/E_0 = 1$ ($T_0 \gg T_b$).}
\label{fig2b}
\end{figure}

A. High-temperature limit ($k_B T_b \gg E_0$). 
The characteristic thermalization time, which in this case we will 
designate by $\tau_H$, is given by 
\begin{eqnarray}
&&{1 \over \tau_H } = {C_{2D} \over \tau_{sc}} \ , \ \ \ \ \ \mbox{where} 
\nonumber \\
&&
C_{2D} = 2 \pi \int_1^{\infty} d \varepsilon \ 
\varepsilon \sqrt{\varepsilon \over \varepsilon - 1} \ 
\Big|F_z \Big(a \sqrt{\varepsilon(\varepsilon - 1)} \Big) \Big|^2 \ . 
\eqnum{15} 
\label{exp4} 
\end{eqnarray} 
In the high-temperature limit of a classical gas of QW excitons 
the thermalization time $\tau_H$ is independent of the bath 
temperature $T_b$. The constant $C_{2D} = C_{2D}(a)$, which completely 
determines $\tau_H$ in terms of $\tau_{sc}$, is much larger than unity. 
For example, for GaAs QW with $L_z = 100 \ \AA$ the dimensionless 
parameter $a = 0.096$ and the constant $C_{2D} = 2530$. 
The corresponding thermalization time of the QW excitons is $\tau_H = 
5.5 \ ps$. For comparison, the high-temperature thermalization time 
of a classical gas of excitons in a bulk semiconductor is given by 
$\tau^{(3D)}_H = (3/8 \pi) \tau_{sc}$ \cite{Ivanov2}. This estimate 
yields $\tau^{(3D)}_H = 1.66 \ ns$ for bulk GaAs, i.e., $\tau^{(3D)}_H 
\gg \tau^{(2D)}_H = \tau_H$. The ratio $\tau^{(2D)}_H / \tau^{(3D)}_H 
= (8 \pi)/(3 C_{2D}) \simeq 3.3 \times 10^{-3}$, which refers to bulk 
GaAs and GaAs QW with $L_z = 100 \ \AA$, clearly demonstrates effective 
cooling of QW excitons in the presence of a bath of bulk phonons. 
Due to the relaxation of momentum conservation in $z$-direction a 
ground-state QW exciton couples to the continuum states $E \geq E_0$ 
rather than to the single energy state $E = E_0$ as occurs in bulk 
materials. With decreasing bath temperature $T_b$ the effective 
thermalization time $\tau_{th}$ of QW excitons monotonously increases 
starting from its high-temperature limit given by $\tau_H$ (see Fig.~2). 
For $T_b = 5 \ K$ one finds from Eq.~(\ref{exp4}) that $\tau_{th} = 
19.1 \ ps$ in GaAs QW with $L_z = 100 \ \AA$. 

There is an uncertainty in values of the deformation potential $D$ of 
exciton -- LA-phonon interaction, published in literature (see, e.g., 
Refs.~\cite{Koteles,Takagahara,Bockelmann}), from $D_{min} = 7.0 \ eV$ 
to $D_{max} = 18.1 \ eV$. In our numerical evaluations we use $D = 15.5 
\ eV$. According to Eq.~(\ref{exp2}), the deformation potential 
contributes to $\tau_{th}$ only through $\tau_{sc} \propto 1/D^2$. 
Therefore the numerical calculations of $\tau_{ph}$ can be 
straightforwardly re-scaled to another value of $D$ (we will use this 
procedure in calculations of Fig.~9a to reproduce qualitatively the 
experimental data of Ref.~\cite{Damen}). The value $D = 15.5 \ eV$ 
corresponds to $\tau_{sc} = 13.9 \ ns$. 

B. Low-temperature limit ($k_BT_b \leq E_0$). In this case 
Eq.~(\ref{exp3}) reduces to 
\begin{equation}
{1 \over \tau_L} = { 2 \pi^{3/2} \over 
\tau_{sc} } \ \left( {E_0 \over k_B T_b} \right)^{1/2}  
e^{-E_0/k_BT_b} \ ,  
\eqnum{16}
\label{exp5} 
\end{equation}
where we designate the low-temperature limit of $\tau_{th}$ by 
$\tau_L$. According to Eq.~(\ref{exp5}) the thermalization time 
$\tau_L$ increases exponentially with decreasing bath temperature $T_b$ 
below $E_0/k_B$. The origin of this result can be understood from the 
initial Eq.~(\ref{intr3}) of the phonon-assisted relaxation kinetics 
into the ground-state mode $E=0$. For a classical gas of QW excitons 
one has $N_E \ll 1$. As a result, both the spontaneous (independent 
of $N_{E=0}$) and the stimulated (proportional to $N_{E=0}$) processes 
contribute to population of the ground-state mode. The first, Stokes, 
process is $\propto N_{E \geq E_0}(1 + n^{ph}_{E \geq E_0})$ and 
increases $N_{E=0}$, while the second, anti-Stokes, one is $\propto - 
N_{E=0} n^{ph}_{E \geq E_0}$ and decreases the occupation number of 
the mode $E=0$. For the high-temperature limit of a classical gas 
of QW excitons both opposite fluxes are intense because 
$n^{ph}_{E \geq E_0} = k_BT_b/E \gg 1$. In the low-temperature limit 
$n^{ph}_{E \geq E_0} = \exp(- E/k_BT_b) \ll 1$ and both the spontaneous 
process, which is proportional to $N_{E \geq E_0}(1 + n^{ph}_{E \geq 
E_0}) \simeq N_{E \geq E_0} = (T_0/T) \exp(-E/k_BT)$, and the 
stimulated process, which is proportional to $- N_{E=0} n^{ph}_{E 
\geq E_0} = - (T_0/T) \exp(-E/k_BT_b)$, are exponentially weak. 

The plot $\tau_{th} = \tau_{th}(T_b)$ (see Fig.~2) shows that for 
excitons in GaAs QWs the high-temperature limit given by 
Eq.~(\ref{exp4}) means $T_b \geq 30 - 50 \ K$. On the other hand, a 
strong increase of $\tau_{th}$ occurs already at $T_b \simeq 
5 - 10 \ K$, i.e., at temperatures much above the low-temperature 
range determined by $T_b \leq E_0/k_b \simeq 0.54 \ K$ and where the 
approximation of  $\tau_{th}$ by Eq.~(\ref{exp5}) is valid. 

2. {\it Quantum gas of QW excitons ($T_0 \gg T_b$).} 

In this case Eq.~(\ref{exp2}) reduces to 
\begin{eqnarray} 
{1 \over \tau_{th} } & = & \left( {2 \pi \over \tau_{sc}} \right) 
\left( {E_0 \over k_B T_0} \right) \int_1^{\infty} d \varepsilon \ 
\varepsilon^2 \sqrt{\varepsilon \over \varepsilon - 1} 
\nonumber \\ 
& \times & 
\Big|F_z \Big(a \sqrt{\varepsilon(\varepsilon - 1)} \Big)\Big|^2 
{ e^{\varepsilon E_0 / k_B T_b} 
\over ( e^{\varepsilon E_0 / k_B T_b} - 1 )^2 } \ . 
\eqnum{17} 
\label{exp6} 
\end{eqnarray} 
Equation~(\ref{exp6}) shows that the thermalization time of quantum 
degenerate quasi-2D excitons depends on the concentration $\rho_{2D}$, 
i.e., $\tau_{th} \propto \rho_{2D}$. 

A. High-temperature limit ($k_B T_b \gg E_0$). In this limit 
Eq.~(\ref{exp6}) yields for $\tau_{0H} = \tau_{th}$:
\begin{eqnarray}
&&{1 \over \tau_{0H} } = \Bigg( {k_B T_b^2 \over T_0 E_0} \Bigg) 
{ {\widetilde C}_{2D} \over \tau_{sc}} \ , 
\ \ \ \mbox{where} 
\nonumber \\
&& {\widetilde C}_{2D} = 2 \pi \int_1^{\infty} d \varepsilon \ 
\sqrt{\varepsilon \over \varepsilon - 1} \ 
\Big|F_z \Big(a \sqrt{\varepsilon(\varepsilon - 1)} \Big) \Big|^2 \ . 
\eqnum{18} 
\label{exp7} 
\end{eqnarray} 
From Eqs.~(\ref{exp4}) and (\ref{exp7}) we find 
\begin{equation}
{1 \over \tau_{0H} } = \Bigg( {k_B T_b^2 \over T_0 E_0} \Bigg) 
\Bigg( { {\widetilde C}_{2D} \over C_{2D} } \Bigg) \ 
{1 \over \tau_H } \ . 
\eqnum{19} 
\label{exp8} 
\end{equation} 
The constant ${\widetilde C}_{2D} = {\widetilde C}_{2D}(a)$ is much 
larger than unity, but much less than $C_{2D}$. For example, 
${\widetilde C}_{2D} = 170$ so that ${\widetilde C}_{2D} / C_{2D} = 
0.07$ for GaAs QW with $L_z = 100 \ \AA$. According to Eq.~(\ref{exp8}) 
one has $\tau_{0H} < \tau_H$ for $T_0 \gg T_b > [(T_0 E_0 C_{2D}) / 
(k_B {\tilde C}_{2D}) ]^{1/2}$ and $\tau_{0H} > \tau_H$ for $T_b < 
[(T_0 E_0 C_{2D}) / (k_B {\tilde C}_{2D}) ]^{1/2}$. The acceleration or 
slowing down of the relaxation kinetics in comparison with that in 
a classical gas of QW excitons originates from BE statistics. 
Because for the quasi-equilibrium degenerate QW excitons $N_{E=0} = 
\exp(T_0/T) \gg N_{E \geq E_0} = 1/[\exp(E_0/k_B T) - 1]$, only the 
stimulated processes $\propto N_{E=0} (N_{E \geq E_0} - n^{ph}_{E \geq 
E_0})$ contribute to occupation kinetics of the ground-state mode $E=0$. 
From Eqs.~(\ref{term4}) and (\ref{term5}) one obtains $\delta T / T = 
- \delta N_{E=0} / N_{E=0}$ for classical statistics and $\delta T / T 
= - (T/T_0) (\delta N_{E=0} / N_{E=0})$ for well-developed BE 
statistics of the ground-state mode, i.e., the same relative change 
of the occupation number $N_{E=0}$ is accompanied at $T \ll T_0$ 
by the much less relative change of the effective temperature than 
that at $T \gg T_0$. Therefore, BE occupation of the ground-state 
mode with $N_{E=0} \gg 1$ slows down the thermalization kinetics of 
QW excitons. On the other hand, one has $N_{E \geq E_0} - n^{ph}_{E 
\geq E_0} = n^{ph}_{E \geq E_0} (\delta T / T_b)$ for well-developed 
BE statistics of the energy modes $E \geq E_0$ and $N_{E \geq E_0} - 
n^{ph}_{E \geq E_0} = n^{ph}_{E \geq E_0} (E / k_BT_b) (\delta T / 
T_b)$ for a classical distribution of QW excitons at $E \geq E_0$. 
As a result, the BE occupation numbers $N_{E \geq E_0} \gg 1$ enhance 
the relaxation dynamics by the factor $\sim k_BT_b/E_0$, which is much 
larger than unity in the high-temperature limit. With decreasing bath 
temperature $T_b$ the slowing down of thermalization, which results 
from $N_{E=0} \gg 1$, starts to dominate over the acceleration of 
relaxation due to BE statistics of the modes $E \geq E_0$. 
\begin{figure}[t]
\begin{center}
\leavevmode
\epsfxsize=8cm \epsfbox{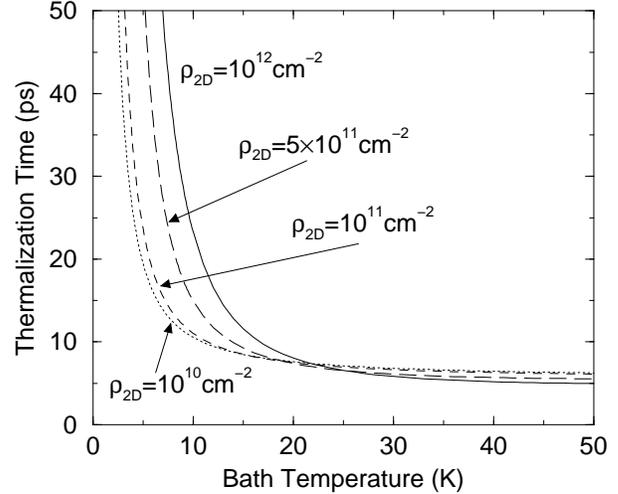}
\end{center}
\caption{ The thermalization time $\tau_{th} = \tau_{th}(T_b)$ of 
BE-distributed QW excitons of the densities $10^{12} \ cm^{-2}$ (solid 
line), $5 \times 10^{11} \ cm^{-2}$ (long dashed line), $10^{11} \ 
cm^{-2}$ (dashed line), and $10^{10} \ cm^{-2}$ (dotted line). The 
calculations with Eq.~(\ref{exp2}) refer to GaAs QW with $L_z = 100 \ 
\AA$ and $\tau_{sc} = 13.9 \ ns$. }
\label{fig3b}
\end{figure}

The thermalization time $\tau_{th} = \tau_{th}(a)$ is very sensitive 
to the dimensionless parameter $a \propto L_z$ through the form-factor 
function $F_z$ on the r.h.s. of Eq.~(\ref{exp2}). With decreasing 
$L_z$ the spectral width of $F_z(a \sqrt{\varepsilon(\varepsilon - 
1)})$ increases indicating a stronger relaxation of the momentum 
conservation in the $z$-direction. From Eqs.~(\ref{exp4}) and 
(\ref{exp7}) we conclude that for the infinite square QW confinement 
potential $\tau_H \propto L_z^2$ and $\tau_{0H} \propto L_z$, 
respectively. The dependences $\tau_H = \tau_H(L_z)$ and $\tau_{0H} = 
\tau_{0H}(L_z)$ are plotted in the inset of Fig.~2. 

B. Low-temperature limit ($k_BT_b < E_0$). In this case we get 
from Eq.~(\ref{exp6}):   
\begin{equation}
{1 \over \tau_{0L}} = { 2 \pi^{3/2} \over 
\tau_{sc} } \ \Bigg( {E_0 T_b \over k_B T_0^2} \Bigg)^{1/2}  
e^{-E_0/k_B T_b} = \Bigg( {T_b \over T_0} \Bigg) \  
{1 \over \tau_L} \ ,  
\eqnum{20}
\label{exp9} 
\end{equation}
where we designate the low-temperature thermalization time of 
degenerate QW excitons by $\tau_{0L}$ and the corresponding 
thermalization time of Maxwell-Boltzmann distributed QW excitons 
$\tau_L$ is given by Eq.~(\ref{exp5}). The slowing down of the 
thermalization process by the factor $T_0/T_b \propto \rho_{2D}$, 
which is much larger than unity, stems from well-developed BE 
statistics of the ground state mode, i.e., from $N_{E=0} \gg 1$. 
The distribution function of QW excitons at $E \geq E_0$ is classical, 
i.e., $N_{E \geq E_0} = \exp(-E_0/k_BT) < 1$, and does not enhance the 
relaxation processes into the ground-state mode $E = 0$. 
\begin{figure}[t]
\begin{center}
\leavevmode
\epsfxsize=8cm \epsfbox{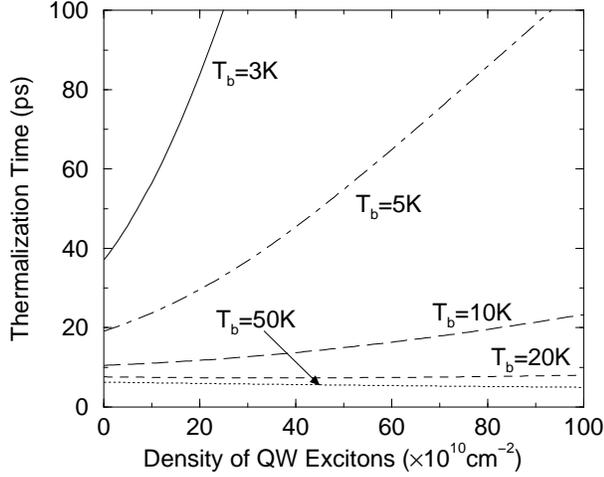}
\end{center}
\caption{ The thermalization time $\tau_{th} = \tau_{th}(\rho_{2D})$ 
of BE-distributed QW excitons at the bath temperatures  $3 \ K$ (solid 
line), $5 \ K$ (dot-dashed line), $10 \ K$ (long dashed line), $20 \ K$ 
(dashed line), and $50 \ K$ (dotted line). The calculations with 
Eq.~(\ref{exp2}) deal with GaAs QW with $L_z = 100 \ \AA$ and $\tau_{sc} 
= 13.9 \ ns$. Note that numerical evaluations of Figs.~2-4 refer to 
effective temperatures of QW excitons close to the bath temperature, 
i.e., $|T - T_b| \ll T_b$. }
\label{fig4b}
\end{figure}

The influence of BE statistics on the thermalization of QW excitons 
is demonstrated in Fig.~3 for $\tau_{th} = \tau_{th}(T_b)$, $\rho_{2D} 
= const.$, and in Fig.~4 for $\tau_{th} = \tau_{th}(\rho_{2D})$, $T_b 
= const.$, respectively. The numerical evaluations of $\tau_{th}$ refer 
to Eq.~(\ref{exp2}). 
\subsection{Thermalization of QW excitons at T$_b$ = 0}
The linearization of the basic Eq.~(\ref{term6}) at $T_b > 0$, 
developed in previous subsection, does not hold when $|T - T_b| > 
k_BT_b^2/E_0$ and especially at zero bath temperature. Because 
$N^{eq}_{E=0} \rightarrow \infty$ at $T_b=0$, the effective temperature 
$T$ of QW excitons is a nonanalytic point of the r.h.s. of 
Eq.~(\ref{term6}) at $T=T_b=0$. For $T_b=0$ Eq.~(\ref{term6}) reduces to 
\begin{eqnarray}
{\partial \over \partial t} T & = & - { 2 \pi \over \tau_{sc} } 
\left( {T^2 \over T_0} \right) \left(1 - e^{-T_0/T} \right) 
\int_1^{\infty} d \varepsilon \ \varepsilon \sqrt{\varepsilon 
\over \varepsilon - 1} 
\nonumber \\
& \times & { \Big|F_z \Big(a 
\sqrt{\varepsilon(\varepsilon - 1)} \Big)\Big|^2 \over 
\Big( e^{\varepsilon E_0/k_BT} + e^{-T_0/T} - 1 \Big) }  \ .  
\eqnum{21}
\label{ground1} 
\end{eqnarray}
Equation (\ref{ground1}), which describes how the QW excitons with 
effective temperature $T$ cool down towards $T_b = 0$, can be further 
simplified for $k_B T \leq E_0$ : 
\begin{equation}
{\partial \over \partial t} T = - { 2 \pi^{3/2} \over 
\tau_{sc} } \ \left( {T^2 \over T_0} \right) 
\left( { k_B T \over E_0 } \right)^{1/2} e^{-E_0/k_BT} \ . 
\eqnum{22}
\label{ground2} 
\end{equation}
In dimensionless time and temperature units $\tau = (2 \pi^{3/2} 
E_0 t)/(k_B T_0 \tau_{sc})$ and ${\tilde T} = k_B T/E_0$ 
Eq.~(\ref{ground2}) takes the canonical form $d {\tilde T} / d \tau 
= - {\tilde T}^{5/2} \exp(-1/{\tilde T})$. The solution $T = T(t)$ 
of Eq.~(\ref{ground2}) is given by the transcendental equation: 
\begin{equation}
F(k_BT/E_0) = { 2 \pi^{3/2} \over \tau_{sc} } \ 
\left( { E_0 \over k_B T_0 } \right) t + A_i \ , 
\eqnum{23}
\label{ground3} 
\end{equation}
where $F(x) = [ 1/\sqrt{x} - Ds(1/\sqrt{x})] e^{1/x}$, $Ds(y) = e^{-y^2} 
{\displaystyle \int}_0^y dt \ e^{t^2}$ is Dawson's integral, and 
the integration constant $A_i$ is defined by the initial condition $T_i 
= T(t=0) \leq E_0/k_B$, i.e., $A_i =F(k_BT_i/E_0)$. For $t \gg [(k_B 
T_0) /(2 \pi^{3/2} E_0)] \tau_{sc}$ the asymptotic solution of 
Eq.~(\ref{ground2}) is 
\begin{equation}
k_B T(t) = { E_0 \over 
\ln \Big[ (2 \pi^{3/2} E_0 t) / (k_B T_0 \tau_{sc}) \Big] } \ , 
\eqnum{24}
\label{ground4} 
\end{equation}
i.e., ${\tilde T}(\tau) = 1/\ln(\tau)$ for $\tau \gg 1$.  

The asymptotic law Eq.~(\ref{ground4}) characterizes the 
{\it nonexponential} and extremely slow thermalization kinetics of a 
quantum gas of QW excitons. Because the phonon occupation numbers 
$n^{ph}_{\bf q} =0$ at $T_b = 0$, only Stokes scattering of the QW 
excitons determines the relaxation process. In this case the 
integrand on the r.h.s. of Eq.~(\ref{intr3}) is proportional to 
$N_{E \geq E_0} (N_{E=0} + 1) \simeq N_{E \geq E_0} N_{E=0}$. At $k_BT 
\leq E_0$ the energy states $E \geq E_0$, which couple with the 
ground-state mode $E=0$ [see Eq.~(\ref{intr3})], are weakly populated 
by the QW excitons, while $N_{E=0} \gg 1$. The Maxwell-Boltzmann 
distribution function $N_{E \geq E_0} = \exp(-E/k_B T)$ gives rise to 
the factor $\exp(-E_0/k_B T) \ll 1$ on the r.h.s. of 
Eq.~(\ref{ground2}). This term together with the need to accumulate a 
huge number of QW excitons in the ground-state mode ${\bf p_{\|}}=0$, 
due to BE statistics, are responsible for $1/\ln(\tau)$ thermalization 
law at $T_b = 0$. 

In Fig.~5 we plot the results of numerical evaluation of the 
relaxational dynamics at $T_b = 0$ from Eq.~(\ref{ground1}) for 
various concentrations $\rho_{2D}$ of QW excitons and for an initial 
effective temperature $T_i = 100 \ K$. The simulations refer to the 
long-lived quasi-2D excitons in GaAs/AlGaAs CQWs. The first hot 
relaxation (see the inset of Fig.~5) is completed within $\Delta t \leq 
30 - 100 \ ps$. At the end of the transient stage the effective 
temperature $T$ of QW excitons is still much higher than $E_0/k_B \simeq 
0.54 \ K$ so that $N_{E \simeq E_0} > 1$ for $T \leq T_0$. As a result, 
the duration of hot thermalization decreases with increasing 
concentration $\rho_{2D}$ of QW excitons. Here one has an acceleration 
of the relaxation kinetics due to BE statistics of energy modes $E \geq 
E_0$. The further thermalization kinetics of cold QW excitons refers to  
$t \geq 100 \ ps$ and reveals $1 / \ln(\tau)$ law (see Fig.~5). 
According to Eq.~(\ref{ground4}), critical slowing down of the 
phonon-assisted relaxation dynamics develops with increasing degeneracy 
temperature $T_0 \propto \rho_{2D}$. An extension of the numerical 
evaluations presented in Fig.~5 on $\mu s$ time scale shows that even 
at $\Delta t = 1 \ \mu s$ the effective temperature of QW excitons is 
still a few hundred $mK$ for $\rho_{2D} \geq 10^{11} \ cm^{-2}$, i.e., 
$T(t= 1\mu s) = 89 \ mK$ for $\rho_{2D} = 10^{10} \ cm^{-2}$, $T(t=1 
\mu s) = 129 \ mK$ for $\rho_{2D} = 10^{11} \ cm^{-2}$, $T(t=1 \mu s) 
= 181 \ mK$ for $\rho_{2D} = 5 \times 10^{11} \ cm^{-2}$, and $T(t=1 
\mu s) = 213 \ mK$ for $\rho_{2D} = 10^{12} \ cm^{-2}$. On a time 
scale much longer than the duration of the initial hot thermalization 
the effective temperature $T$ is nearly independent of $T_i = T(t=0)$, 
provided that $k_BT_i \gg E_0$. Recently, a strong increase of the 
thermalization times in a highly degenerate quantum gas of quasi-2D 
excitons has indeed been observed in GaAs/AlGaAs CQWs \cite{Butov3}. 
\begin{figure}[t]
\begin{center}
\leavevmode
\epsfxsize=8cm \epsfbox{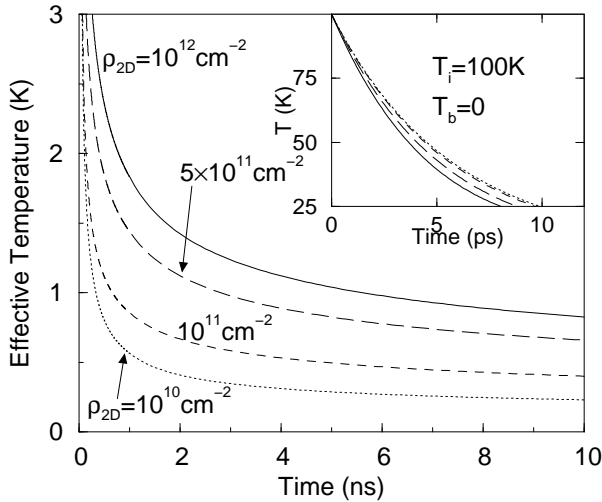}
\end{center}
\caption{ Thermalization dynamics $T = T(t)$ of QW excitons at $T_b = 
0$, $\rho_{2D} = 10^{12} \ cm^{-2}$ (solid line), $5 \times 10^{11} \ 
cm^{-2}$ (long dashed line), $10^{11} \ cm^{-2}$ (dashed line), and 
$10^{10} \ cm^{-2}$ (dotted line). Inset: the initial transient 
thermalization of hot QW excitons BE-distributed at $t=0$ with $T_i = 
100 \ K$. }
\label{fig5b}
\end{figure}

However, our treatment has neglected possible low-temperature 
collective states due to interactions between QW excitons. 
At a critical temperature $T_c = \alpha T_0$, where $\alpha = 
\alpha(a_x^{(2D)}\rho_{2D}^{1/2}) < 1$, a system of quasi-2D excitons 
may undergo a phase transition to a superfluid state. Our calculations 
of relaxational thermodynamics become invalid at $T \leq T_c$, i.e., one 
cannot trace with Eq.~(\ref{ground1}) the transition to the excitonic 
superfluid phase and how further a collective ground state of QW 
excitons arises at $T \rightarrow 0$ (for $\rho_{2D}[a_x^{(2D)}]^2 < 
1$ the ground state can be interpreted in term of BE condensation of 
QW excitons \cite{Littlewood1,Littlewood2}). While in the dilute limit 
$\ln(\rho_{2D}[a_x^{(2D)}]^2) \ll 1$ the parameter $\alpha \ll 1$, i.e., 
$T_c \ll T_0$ \cite{Popov}, there is still no first-principle theory 
to estimate $\alpha$ for the densities $\rho_{2D}[a_x^{(2D)}]^2 
\leq 1$. With the mentioned above restrictions, we find that cooling 
of high-density QW excitons to very low temperatures is rather 
difficult. This conclusion may have some bearing on the search 
\cite{Butov2} for the collective ground state of indirect excitons in 
GaAs/AlGaAs CQWs. 
\begin{figure}[t]
\begin{center}
\leavevmode
\epsfxsize=8cm \epsfbox{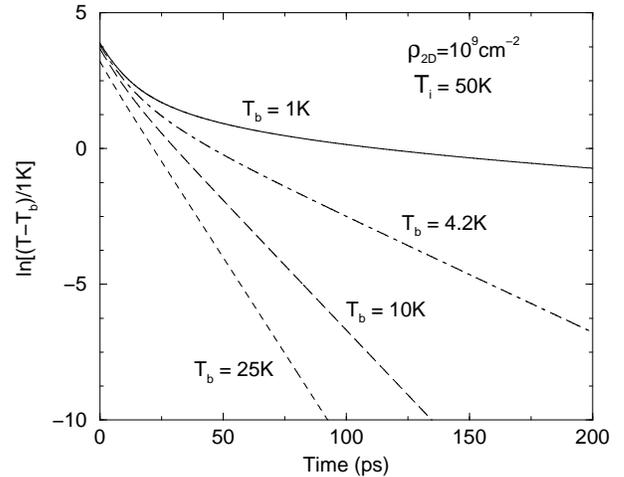}
\end{center}
\caption{ The relaxational dynamics of a classical gas of hot QW 
excitons at $T_b = 1~K$ (solid line), $4.2~K$ (dot-dashed line), 
$10~K$ (long dashed line), and $25~K$ (dashed line). The latter 
straight line refers to the exponential kinetics $\ln|(T - 
T_b)/(T_i - T_b)| = - t/\tau_H$ which is valid in the 
high-temperature limit $k_B T_b \gg E_0$. }
\label{fig6b}
\end{figure}
\subsection{Thermalization dynamics of hot QW excitons 
(T~$\gg$~T$_b$)} 
In experiments [1-9] with nonresonant excitation of heavy-hole 
excitons in GaAs QWs the initial effective temperature $T_i 
\gg T_b$. While in this case the thermalization kinetics can be 
analyzed numerically from the basic Eq.~(\ref{term6}), we will 
clarify the various relaxation scenarios by analytic approximations 
of Eq.~(\ref{term6}). 

1. {\it Classical gas of QW excitons ($T_i \gg T_b \gg T_0$). } 

A. High-temperature limit ($k_B T_b \gg E_0$). In this case 
Eq.~(\ref{term6}) reduces to 
\begin{equation}
{\partial \over \partial t} T = - {1 \over \tau_H} \ 
(T - T_b) \ ,  
\eqnum{25}
\label{hot1} 
\end{equation}
where the characteristic thermalization time $\tau_H$ is given 
by Eq.~(\ref{exp4}). Thus, in the high-temperature limit 
$k_B T_b \gg E_0$ the exponential law $T(t) = T_b + (T_i - 
T_b) \exp( -t/\tau_H )$ is valid even for $(T_i-T_b)/T_b \gg 1$. 

B. Low-temperature limit ($k_BT_b < E_0$). The initial hot 
thermalization down to $k_B T \simeq E_0$ is approximated by 
\begin{equation}
{\partial \over \partial t} T = - {T \over \tau_H} \ . 
\eqnum{26}
\label{hot2} 
\end{equation} 
This exponential cooling $T(t) = T_i \exp( -t/\tau_H)$ completes 
at $t' \simeq \tau_H \ln(T_i/T_b)$. The hot thermalization 
refers to the high-temperature limit given by Eq.~(\ref{hot1}). 
At $t > t'$ the cold QW excitons relax to the bath temperature 
according to 
\begin{equation}
{\partial \over \partial t} T = - {1 \over \tau_L} 
(T - T_b) \ ,  
\eqnum{27}
\label{hot3} 
\end{equation}
where the low-temperature thermalization time $\tau_L$ is defined 
by Eq.~(\ref{exp5}). Approximation (\ref{hot3}) is valid for $(T - 
T_b)/T_b < k_BT_b/E_0$, yields the exponential relaxation with $\tau_L$,  
and corresponds to the linearization of Eq.~(\ref{term6}) at $T=T_b$. 

If the high-temperature limit does not hold, the thermalization 
kinetics of Maxwell-Boltzmann distributed QW excitons is exponential 
only locally in time, i.e., $\tau_{th} = \tau_{th}(t)$. The described 
above picture of the relaxation in the low-temperature limit in fact 
deals with a continuous increase of the instant thermalization time 
$\tau_{th}(t)$ from $\tau_H$ to $\tau_L$. As a result, for $T_i - T_b 
\gg T_b$ the total relaxation process can show a nonexponential 
behavior. In Fig.~6 we illustrate how a single-exponential 
thermalization kinetics with $\tau_{th} = \tau_H$ of a classical gas of 
QW excitons initially distributed at $T_i \gg T_b$ develops with the 
decreasing bath temperature towards the nonexponential relaxation. 
\begin{figure}[t]
\begin{center}
\leavevmode
\epsfxsize=8cm \epsfbox{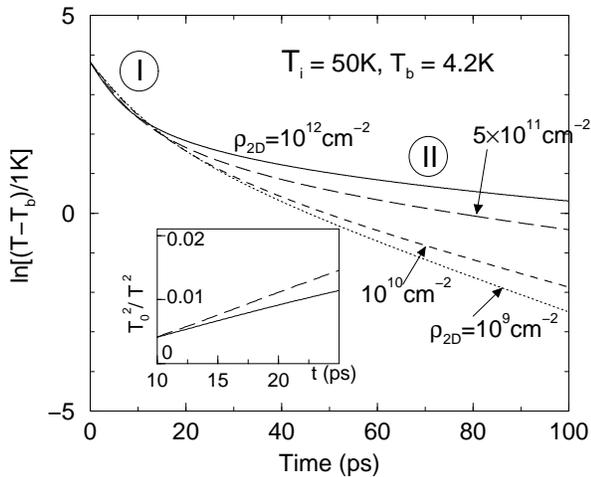}
\end{center}
\caption{ Thermalization kinetics of hot QW excitons of the densities 
$10^{12} \ cm^{-2}$ (solid line), $5 \times 10^{11} \ cm^{-2}$ (long 
dashed line), $10^{10} \ cm^{-2}$ (dashed line), and $10^{9} \ 
cm^{-2}$ (dotted line). The initial effective temperature of QW 
excitons $T_i = 50~K$, the bath temperature $T_b = 4.2~K$. GaAs QW 
with $L_z = 100~\AA$ and $\tau_{sc} = 13.9 \ ns$. The first transient 
thermalization occurs in the time interval $I$ ($t \leq 20 - 40 ps$). 
The density-dependent quasi-steady-state relaxation develops  in the 
time-domain $II$, where $T - T_b < T_b$. Inset: verification of the 
approximation Eq.~(\ref{hot5}) of nonexponential transient relaxation 
of strongly degenerate QW excitons. }
\label{fig7b}
\end{figure}

2. {\it Quantum gas of QW excitons ($T_i \gg T_0 \gg T_b$). } 

A. High-temperature limit ($k_B T_b \gg E_0$). The initial hot 
relaxation to the temperature $T \simeq T_0$ is given by 
Eq.~(\ref{hot2}) and completes at $t' \simeq \tau_H \ln(T_i/T_0)$. 
At $t \geq t'$ QW excitons become degenerate 
and BE statistics starts to influence the thermalization process. 
The relaxation kinetics is approximated by 
\begin{equation}
{\partial \over \partial t} T = - \left( { T \over T_b } 
\right)^2 \ { (T - T_b) \over \tau_{0H}} \ ,  
\eqnum{28}
\label{hot4} 
\end{equation} 
where $\tau_{0H} \propto \rho_{2D}$ is given by Eq.~(\ref{exp7}). 
Equation (\ref{hot4}) yields the nonexponential transient 
thermalization from $T \simeq T_0$ to $T \simeq T_b$:
\begin{equation}
T(t) =  { T_0 \over \Big[1 + 2(T_0/T_b)^2  
[(t-t')/\tau_{0H}] \Big]^{1/2}} \ . 
\eqnum{29}
\label{hot5} 
\end{equation} 
The nonexponential transient lasts for a time $\Delta t \simeq 
\tau_{0H}/2$. The relaxation kinetics from $t \geq t'' \simeq t' 
+ \tau_{0H}/2$ refers to $T - T_b < T_b$ and recovers the exponential 
law with $\tau_{0H}$. 

B. Low-temperature limit ($k_BT_b < E_0$). At $0 \leq t \leq 
t' \simeq \tau_H \ln(T_i/T_0)$ the first hot thermalization 
from $T=T_i$ to $T \simeq T_0$ is characterized by the exponential 
law of Eq.~(\ref{hot2}). The next nonexponential transient 
relaxation from $T \simeq T_0$ to $T \simeq E_0/k_B$ is given by 
Eq.~(\ref{hot5}) and corresponds to the time interval $t' \leq t 
\leq t'' \simeq (k_B T_b/E_0)^2 (\tau_{0H}/2)$. These two stages 
are similar to those found in the high-temperature limit. For the 
thermalization from $T \simeq E_0/k_B$ to $T \simeq T_b$ one 
recovers Eq.~(\ref{ground2}) and the corresponding $1 / \ln(\tau)$ 
kinetics of Eq.~(\ref{ground4}) treated in the previous subsection for 
$T_b = 0$. This stage is rather long and completes at $t''' \simeq 
(k_B T_0 \tau_{sc}/2\pi^{3/2} E_0) \exp(E_0/k_BT_b) = 
(k_BT_b/E_0)^{1/2} \tau_{0L}$. The final exponential relaxation 
refers to $(T - T_b)/T_b < k_BT_b/E_0$ and is given by $T(t \geq 
t''') = T_b + (k_B T^2_b / E_0) \exp[-(t-t''')/\tau_{0L}]$, where 
the low-temperature thermalization time $\tau_{0L}$ of the 
degenerate QW excitons is defined by Eq.~(\ref{exp9}). 

In order to demonstrate how BE statistics can influence the 
thermalization kinetics of QW excitons at $T_i \gg T_b$, in Fig.~7 we 
plot the numerical solution $T=T(t)$ of Eq.~(\ref{term6}) for $T_i = 
50 \ K$, $T_b = 4.2 \ K$, and various values of the concentration 
$\rho_{2D}$. The development of the nonexponential relaxation with 
the increasing degeneracy temperature $T_0 \propto \rho_{2D}$ is 
clearly seen. The insert of Fig.~7 illustrates the transient 
thermalization given by Eq.~(\ref{hot5}). 
\section{Resonant photoluminescence of quantum degenerate 
quasi-2D excitons}
In high-quality GaAs QWs the decay of quasi-2D excitons is mainly due 
to radiative recombination 
\cite{Feldmann,Damen,Deveaud1,Vinattieri,Kuhl,Wood,Weimann,Shah0}. 
Moreover, following the first studies \cite{Golub} of the optical 
properties of excitons in GaAs/AlGaAs CQWs, very recent experiments 
\cite{Butov1} show that the radiative recombination channel can also 
be dominant for the long-lived indirect excitons. In this section we 
generalize the theory \cite{Feldmann,Andreani1} of steady-state 
resonant PL of a quasi-equilibrium classical gas of QW excitons to 
well-developed BE statistics and exploit the relaxation 
thermodynamics of Eq.~(\ref{term6}) in order to analyze the PL 
kinetics of QW excitons. 

In perfect QWs an exciton can emit a bulk photon only from radiative 
modes, which are located inside the photon cone given by $k = k(\omega) 
= (\sqrt{\epsilon_b} \omega )/(\hbar c)$. Here $\omega$ is the 
frequency of light and $\epsilon_b$ is the background dielectric 
constant for an exciton line. This means that the radiative zone of QW 
excitons is given by $p_{\|} \leq k_0$, where $k_0 =  k(\omega_t)$ 
corresponds to crossover of the photon and exciton dispersions and 
$\hbar \omega_t$ is the exciton energy. The intrinsic radiative rates 
for in-plane transverse ($T$-polarized) and in-plane longitudinal 
($L$-polarized) dipole-active QW excitons are given by 
\cite{Andreani1} 
\begin{equation}
\Gamma_T(p_{\|}) = \Gamma_0 { k_0 \over \sqrt{k_0^2 - p_{\|}^2} } 
\ \ \ \mbox{and} \ \ \  
\Gamma_L(p_{\|}) = \Gamma_0 { \sqrt{k_0^2 - p_{\|}^2} \over k_0 } \ , 
\eqnum{30}
\label{pl1} 
\end{equation} 
respectively. The intrinsic radiative lifetime $\tau_R$ of a QW 
exciton is defined by $\tau_R = 1/\Gamma_0$, where $\Gamma_0 
= \Gamma_0(L_z) = 
\Gamma_T(p_{\|}=0) = \Gamma_L(p_{\|}=0)$ is the radiative rate for 
the ground-state mode ${\bf p_{\|} } = 0$. $Z$-polarized heavy-hole QW 
excitons are forbidden. Here we use the polarization notations of 
Ref.~\cite{Andreani1}. 

The total radiative decay rate $\Gamma_{opt} = 1 / \tau_{opt}$ of 
a gas of quasi-equilibrium BE-distributed QW excitons is 
\begin{equation}
\Gamma_{opt} = { 1 \over \rho_{2D} } \ \int_0^{k_0} 
N_{p_{\|}} \Big [\Gamma_T(p_{\|}) + \Gamma_L(p_{\|}) \Big] { p_{\|} 
dp_{\|} \over 2 \pi } \ , 
\eqnum{31}
\label{pl2} 
\end{equation} 
where the occupation number $N_{p_{\|}} = N^{eq}_{E_{p_{\|}}}$ is given 
by Eq.~(\ref{term3}). Equation (\ref{pl2}) takes into account the equal 
probabilities of the $T$- and $L$-polarizations and two-fold spin 
degeneracy, $\sigma^+$ and $\sigma^-$, of dipole-active QW excitons. 
Using Eqs.~(\ref{term3}) and (\ref{pl1}) we find from Eq.~(\ref{pl2}): 
\begin{eqnarray}
\Gamma_{opt} & = & { \Gamma_0 \over 2 } \ 
\Big [J_T(T,T_0) + J_L(T,T_0) \Big]  \ , 
\nonumber \\ 
J_T(T,T_0) & = & \Bigg( { E_{k_0} \over k_B T_0 } \Bigg) \ \int_0^{1} 
{ dz \over A e^{- z^2 E_{k_0}/k_B T} - 1 } \ , 
\nonumber \\ 
J_L(T,T_0) & = & \Bigg( { E_{k_0} \over k_B T_0 } \Bigg) \ \int_0^{1} 
{ z^2 dz \over A e^{- z^2 E_{k_0}/k_B T} - 1 } \ , 
\eqnum{32}
\label{pl3} 
\end{eqnarray} 
where $A = A(T,T_0) = e^{E_{k_0}/k_B T}/(1 - e^{- T_0/T})$ and $E_{k_0} 
= \hbar^2 k_0^2/2M_x$ (for GaAs QWs with $\epsilon_b = 12.9$ and $\hbar 
\omega_t = 1.55~eV$ one has $E_{k_0} \simeq 101 \ \mu eV$ and 
$E_{k_0}/k_B \simeq 1.17 \ K$). The contribution of two-fold 
dipole-inactive (triplet) QW excitons to the total density $\rho_{2D}$ 
is included in Eq.~(\ref{pl3}) through the degeneracy temperature 
$T_0 \propto \rho_{2D}$. 

For Maxwell-Boltzmann distributed QW excitons at the effective 
temperature $T$ much larger than $T_0$ and $E_{k_0}/k_B$ we find from 
Eq.~(\ref{pl3}) that $J_T = 3J_L = (\hbar^2 k_0^2)/(M_x T)$ and 
\begin{equation}
\Gamma^{cl}_{opt} = \Bigg( { \hbar^2 k_0^2 \over 3 M_x k_B T } \Bigg) 
\Gamma_0 \ .
\eqnum{33}
\label{pl4} 
\end{equation} 
In the limit $T \ll T_0$ of a strongly degenerate gas of QW 
excitons one approximates $J_T = 1 + (T/T_0) \ln(4 E_{k_0}/k_b T)$ 
and $J_L = 1 + (T/T_0) \Big[ \ln(4 E_{k_0}/k_b T) - 2 \Big]$. In this 
case $\Gamma_{opt}$ of Eq.~(\ref{pl3}) reduces to 
\begin{equation}
\Gamma^q_{opt} = \Bigg[ 1 - {T \over T_0} + {T \over T_0} 
 \ln \Bigg( { 2 \hbar^2 k_0^2 \over M_x k_B T } \Bigg) \Bigg] 
{\Gamma_0 \over 2} \ .
\eqnum{34}
\label{pl5} 
\end{equation} 
The expansion of $J_L(T,T_0)$, $J_T(T,T_0)$, and $\Gamma_{opt}$ in 
terms of the ratio $T/T_0 \ll 1$ is valid for the effective 
temperature of QW excitons $T \leq E_{k_0} \exp(-T_0/T)$. The second 
and third terms in the square brackets on the r.h.s. of Eq.~(\ref{pl5}) 
are small corrections, i.e., $\Gamma^q_{opt} (T \rightarrow 0) 
\rightarrow \Gamma_0/2$. The nonzero limit of $\Gamma_{opt}$ at $T 
\rightarrow 0$, which is completely determined by the intrinsic 
radiative rate $\Gamma_0$ of the ground-state mode ${\bf p}_{\|} = 0$, 
is due to the effective accumulation of low-energy QW excitons with 
$N_E \gg 1$ in the radiative zone $p_{\|} \leq k_0$. 

In the limit $T \gg T_0, E_{k_0}$ of a classical behavior one derives 
from Eqs.~(\ref{pl2})-(\ref{pl3}): 
\begin{equation}
\tau_{opt}^{cl} = \Bigg({ 3 M_x k_B T \over \hbar^2 k_0^2 } 
\Bigg) \tau_R + \Bigg( {9 \over 10} - { 3M_x k_B T_0 \over 
2 \hbar^2 k_0^2 } \Bigg) \tau_R \ , 
\eqnum{35}
\label{pl6a} 
\end{equation} 
where in the expansion of $\tau_{opt}^{cl} = 1/\Gamma_{opt}$ we keep 
not only the leading term $1/\Gamma^{cl}_{opt} \propto T$ [see 
Eq.~(\ref{pl4})], but also the temperature-independent 
correction. The next terms of the expansion are proportional 
to $1/T^{n \geq 1} \rightarrow 0$ and can indeed be neglected. 
The first term on the r.h.s. of Eq.~(\ref{pl6a}), i.e., 
$1/\Gamma^{cl}_{opt} = (3 M_x k_B T \tau_R)/(\hbar^2 k_0^2)$, is a 
well-known result of Ref.~\cite{Andreani1}. The 
temperature-independent correction, which is given by the second 
term on the r.h.s. of Eq.~(\ref{pl6a}), originates from the BE 
distribution function used in Eq.~(\ref{pl2}) and consists of the 
density-independent and density-dependent contributions of the 
opposite signs. The density-dependent contribution can be much larger 
than $\tau_R$ if $k_B T \gg k_B T_0 \gg E_{k_0}$. In the leading term 
$1/\Gamma^{cl}_{opt} \propto T$ one has that the optical decay time 
$\tau^{cl}_{opt} \gg \tau_R$, because only a small fraction of QW 
excitons occupy the radiative modes $p_{\|} \leq k_0$ at $T \gg T_0, 
E_{k_0}/k_B$ \cite{Feldmann,Andreani1}. 

In Fig.~8 we plot the effective decay time of quasi-equilibrium QW 
excitons $\tau_{opt} = \tau_{opt}(T)$, calculated numerically with 
Eq.~(\ref{pl3}) for various values of the concentration $\rho_{2D}$. 
Following Ref.~\cite{Andreani1}, for a GaAs QW with $L_z = 100 \ \AA$ 
we use $\tau_R = 25 \ ps$. For $T \gg T_0, E_{k_0}/k_B$ the linear 
behavior given by Eq.~(\ref{pl6a}) is clearly seen (dotted, dashed, 
and long dashed lines in Fig.~8). The reference bold solid line 
indicates $1/\Gamma^{cl}_{opt} \propto T$ law. Note that the linear 
asymptotics of $\tau_{opt} = \tau_{opt}(T,\rho_{2D})$ from the 
classical range $T \gg T_0, E_{k_0}/k_B$ down to low temperatures $T 
\rightarrow 0$ reveal nonzero density-dependent values. Thus one 
concludes that even at high temperatures $T \gg T_0$ the influence 
of nonclassical statistics can be found in the PL of QW excitons. 
At $T \rightarrow 0$ the limit $\tau^{cl}_{opt} \rightarrow \tau_R 
[9/10 - (3k_BT_0)/(4E_{k_0})]$ given by Eq.~(\ref{pl6a}) breaks and 
all curves approach the double intrinsic radiative lifetime $2 
\tau_R$. The density-dependent deviation from the linear classical 
law of Eq.~(\ref{pl6a}) develops with increasing $\rho_{2D}$. For 
example, at $\rho_{2D} = 5 \times 10^{11} \ cm^{-2}$ the quantum 
asymptotics $\tau^q_{opt}(T \rightarrow 0) = 2 \tau_R$ can be traced 
already for the effective temperature $T \simeq 10 \ K$ (see Fig.~8). 
\begin{figure}[t]
\begin{center}
\leavevmode
\epsfxsize=8cm \epsfbox{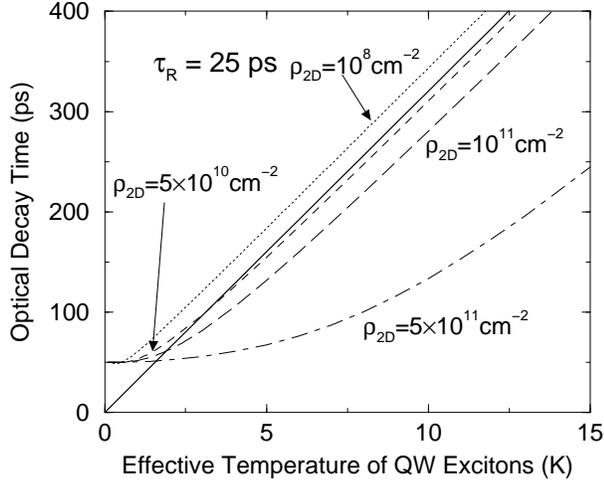}
\end{center}
\caption{ The optical decay time $\tau_{opt}$ versus the effective 
temperature $T$ of QW excitons: $\rho_{2D} = 10^8 \ cm^{-2}$ 
(dotted line), $5 \times 10^{10} \ cm^{-2}$ (dashed line), 
$10^{11} \ cm^{-2}$ (dot-dashed line), and $5 \times 
10^{11} \ cm^{-2}$ (solid line). The intrinsic radiative 
lifetime of QW excitons is $\tau_R = 25 \ ps$. The bold straight line 
is given by $1/\Gamma_{opt}^{cl} = (3M_xk_BT)/(\hbar^2 k_0^2)$. }
\label{fig8b}
\end{figure}

According to Eq.~(\ref{pl1}), the decay rate of $T$-polarized QW 
excitons diverges at $p_{\|} \rightarrow k_0$ \cite{Andreani1}. 
In Appendix~B we show how the homogeneous linewidth of low-energy 
QW excitons $\hbar \Gamma_{hom} = \hbar/\tau_{x-x} + \hbar/\tau_{ph} + 
\hbar/\tau_R \simeq \hbar/\tau_{x-x}$ removes this divergence [see 
Eq.~(B6)]. In the presence of homogeneous broadening, due to dominant 
exciton-exciton scattering, $\Gamma_T(p_{\|})$ of Eq.~(\ref{pl1}) 
changes only in a very narrow band of QW states given by $0 \leq k_0 - 
p_{\|} \leq {\tilde \gamma}k_0$, where ${\tilde \gamma} = 1/(\tau_{x-x} 
\omega_t) \ll 1$. In particular, $\Gamma_T(p_{\|} \rightarrow k_0) = 
\Gamma_0/(2 {\tilde \gamma}^{1/2})$ rather than diverges. Furthermore, 
homogeneous broadening of QW excitons from the radiative zone only 
slightly renormalizes the PL efficiency: $\Gamma_{opt}^{cl}$ changes 
on $\Gamma_{opt}^{cl}[1 - (3/4){\tilde \gamma}^{1/2}]$ (for details 
see Appendix~B). The correction is even less for $\Gamma_{opt}^q$ of 
Eq.~(\ref{pl5}). Therefore we conclude that the corrections to the 
optical decay rate $\Gamma_{opt}$ due to the homogeneous linewidth 
$\Gamma_{hom}$ can indeed be neglected. Note, however, that 
inhomogeneous broadening strongly influences the PL process and 
changes our results obtained for a perfect QW. 

The optical decay of QW excitons from the radiative zone $p_{\|} \leq 
k_0$ does not violate the assumptions of relaxational thermodynamics 
provided that $\tau_{x-x} \ll \tau_R$. In this case the thermal 
quasi-equilibrium distribution of the finite lifetime QW excitons 
at $E \leq E_{k_0}$ holds due to exciton-exciton scattering. Because 
both the minimal thermalization time $\tau_{th}$ given by $\tau_H$ 
of Eq.~(\ref{exp4}) and the intrinsic radiative time $\tau_R$ are on 
the same time scale of $5 - 50 \ ps$, the condition $\tau_{x-x} \ll 
\tau_H$ [see Eq.~(\ref{term7})] of the thermodynamic picture also 
nearly guarantees $\tau_{x-x} \ll \tau_R$. 

The radiative decay leads to the continuous decrease of the density of 
QW excitons (or the degeneracy temperature $T_0 \propto \rho_{2D}$) and 
gives rise to the PL signal according to 
\begin{eqnarray}
{\partial \over \partial t} \rho_{2D} & = & - \Gamma_{opt}(T_0,T) 
\rho_{2D} \equiv - { \rho_{2D} \over \tau_{opt}(\rho_{2D},T) } \ , 
\eqnum{36a} \label{pl6aa} \\
I_{PL} & = & - \hbar \omega_t {\partial  \rho_{2D} \over \partial t}
\equiv \hbar \omega_t { \rho_{2D} \over \tau_{opt}(\rho_{2D},T) } \ , 
\eqnum{36b}
\label{pl6} 
\end{eqnarray} 
where $I_{PL} = I_{PL}(t)$ is the photoluminescence intensity. Thus 
Eqs.~(\ref{term6}), (\ref{pl3}), and (\ref{pl6aa})-(\ref{pl6}) describe 
the PL kinetics of QW excitons cooling from the initial effective 
temperature $T_i$ to the bath $T_b$. For a high-temperature classical 
gas of QW excitons the optical decay time $\tau_{opt}^{cl}$ of 
Eq.~(\ref{pl6a}) is approximated by the density-independent 
$1/\Gamma_{opt}^{cl}$ of Eq.~(\ref{pl4}) and the solution of 
Eq.~(\ref{pl6aa}) is simply $\rho_{2D}(t) = \rho_{2D}^{(0)} 
\exp(- \int_0^t \Gamma_{opt}^{cl}(t) dt)$, where $\rho_{2D}^{(0)}$ 
is the initial concentration of QW excitons and $\Gamma_{opt}^{cl}$ is 
time-dependent through the effective temperature of QW excitons $T = 
T(t)$. In the general case, however, Eq.~(\ref{pl6aa}) gives a 
nonexponential decay of the density. The characteristic times of the 
both fundamental processes which contribute to the PL kinetics, 
LA-phonon-assisted thermalization and optical decay, generally depend 
upon the effective temperature and density of QW excitons, i.e., 
$\tau_{th} = \tau_{th}(T,T_0)$ and $\tau_{opt} =\tau_{opt}(T,T_0)$. 
As a result, the PL kinetics given by Eqs.~(\ref{term6}), (\ref{pl3}), 
and (\ref{pl6aa})-(\ref{pl6}) is also $T$- and $\rho_{2D}$-dependent. 
The thermalization process and radiative decay work in the opposite 
directions with respect to the quantum-statistical effects. Cooling 
of QW excitons is accompanied by an increasing number of low-energy 
particles towards well-developed BE statistics. In the meantime 
$\tau_{opt}$ decreases and the optical decay of $\rho_{2D}$ speeds up 
resulting in decreasing $T_0$. Thus the radiative processes interfere 
with the development of BE statistics. 

For a classical gas of QW excitons in the high-temperature limit $k_B 
T_b \gg E_0$ the PL dynamics of Eqs.~(\ref{term6}), (\ref{pl3}), and 
(\ref{pl6aa})-(\ref{pl6}) can be analyzed analytically. In this case, 
$T(t) = T_b + (T_i - T_b) \exp(-t/\tau_H)$, where the 
temperature-independent $\tau_H$ is given by Eq.~(\ref{exp4}). At the 
first transient thermalization of hot QW excitons from the initial 
distribution at effective $T_i \gg T_b$ one has that 
$\tau^{cl}_{opt}(T_i) \gg \tau_H$, i.e., the optical decay is very slow 
and practically does not change the concentration of QW excitons. The 
transient stage lasts a few $\tau_H$ and for this time domain 
Eq.~(\ref{pl6}) reduces to 
\begin{eqnarray}
I_{PL}(t \leq t_{tr}) & = & 
\hbar \omega_t \Bigg( { \hbar^2 k_0^2 \over 3 M_x k_B T_i } 
\Bigg) \rho_{2D}^{(0)} e^{t/\tau_H} 
\nonumber \\ 
& \equiv & \hbar \omega_t \ 
{ \rho_{2D}^{(0)} \over \tau_{opt}^{cl}(T_i) } \ e^{t/\tau_H} \ , 
\eqnum{37}
\label{pl8} 
\end{eqnarray}
i.e., $I_{PL}(t \leq t_{tr}) \propto \exp(t/\tau_H)$. Therefore at the 
hot relaxation stage $t \leq t_{tr}$ the PL intensity exponentially 
increases with the rate $1/\tau_H$, due to the population of the 
radiative zone. The transient stage completes when $T(t_{tr}) - T_b 
\leq T_b$. The PL intensity gets its maximum at $t = t_0 > t_{tr}$, when 
the population efficiency of the modes $p_{\|} \leq k_0$ is compensated 
by the increasing optical decay: 
\begin{equation}
t_0 = \tau_H \ln \Bigg( {3 M_x k_B T_i \over \hbar^2 k_0^2} {\tau_R 
\over \tau_H} \Bigg) \equiv \tau_H \ln \Bigg[ {\tau^{cl}_{opt}(T_i) 
\over \tau_H} \Bigg] \ . 
\eqnum{38}
\label{pl9} 
\end{equation}
The corresponding effective temperature of QW excitons is $T(t_0) = 
T_b[1 + \tau_H/\tau^{cl}_{opt}(T_i)]$, i.e., $T(t_0) - T_b \ll T_b$. At 
$t = t_s > t_0$ the PL dynamics becomes steady-state and is determined 
by the optical decay, because the QW excitons are already thermalized at 
$T = T_b$. For this time domain the PL kinetics is given by 
\begin{eqnarray}
I_{PL}(t \geq t_s) & = & \hbar \omega_t \Bigg( { \hbar^2 k_0^2 \over 
3 M_x k_B T_b } \Bigg) \rho_{2D}^{(s)} e^{-t/\tau^{cl}_{opt}(T_b)} 
\nonumber \\
& \equiv & \hbar \omega_t \ { \rho_{2D}^{(s)} \over 
\tau_{opt}^{cl}(T_b) } \ e^{-t/\tau^{cl}_{opt}(T_b)} \ ,  
\eqnum{39}
\label{pl10} 
\end{eqnarray}
where $\rho_{2D}^{(s)} = \rho_{2D}(t_s) \simeq \rho_{2D}^{(0)}$. Thus at 
the steady-state regime $I_{PL}(t \geq t_s) \propto 
\exp[-t/\tau_{opt}^{cl}(T_b)]$, i.e., the PL intensity decreases 
exponentially with the optical decay rate $\Gamma_{opt}^{cl}(T_b)$. 
Note that because usually $T_i \gg T_0$ and $T_0(t \rightarrow \infty) 
\rightarrow 0$, the very first transient stage and the very last 
steady-state decay of excitonic PL always follow Eqs.~(\ref{pl8}) and 
(\ref{pl10}), respectively. 

In Fig.~9a we plot numerical simulations of the PL dynamics done with 
Eqs.~(\ref{term6}), (\ref{pl3}), and (\ref{pl6aa})-(\ref{pl6}) for 
excitons of various initial densities in GaAs QW with $L_z = 100 ~\AA$. 
In order to model the experimental data on the density-dependent PL 
kinetics plotted in Fig.~2 of Ref.~[4a] we use $\tau_{sc} = 85~ns$ and 
$\tau_R = 40~ps$. The change of the PL dynamics with increasing 
$\rho_{2D}^{(0)}$ shown in Fig.~9a reproduces qualitatively the 
corresponding experimental observations of Ref.~[4a]. The decrease of 
the rise time $t_0$ of the PL signal with the increasing initial 
density of QW excitons is mainly due to increase of the optical decay 
rate at high densities (see the inset of Fig.~9a). This effect 
originates from BE statistics of QW excitons. 

In Fig.~9b we show numerical modeling of PL kinetics in GaAs/AlGaAs 
CQWs with the intrinsic radiative lifetime of long-lived excitons 
$\tau_R = 30~ns$. Here BE statistics strongly influences the PL temporal 
behavior in the steady-state regime at $t \geq t_s$, but still before a 
time when indirect excitons become Maxwell-Boltzmann distributed due to 
continuous decrease of $\rho_{2D}$. At this time domain the 
thermalization kinetics of high-density CQW excitons undergoes critical 
slowing down, the effective temperature of CQW excitons is nearly 
stabilized at $T_0 > T > T_b$, and the occupation numbers of the 
radiative modes $N_{p_{\|} \leq k_0} \gg 1$. As a result, with the 
increasing initial concentration $\rho_{2D}^{(0)}$ the decay rate of 
the PL signal approaches $\tau_{opt}^q = 2 \tau_R$, i.e., one has an 
acceleration of the PL optical decay owing to well-developed BE 
statistics [see Fig.~9b, where the long dashed and solid reference lines 
indicate $\ln(I_{PL}/I_0) = (t_0 - t)/2\tau_{opt}(T_b)$ and 
$\ln(I_{PL}/I_0) = (t_0 - t)/2\tau_R$, respectively].  
\begin{figure}[t]
\begin{center}
\leavevmode
\epsfxsize=8cm \epsfbox{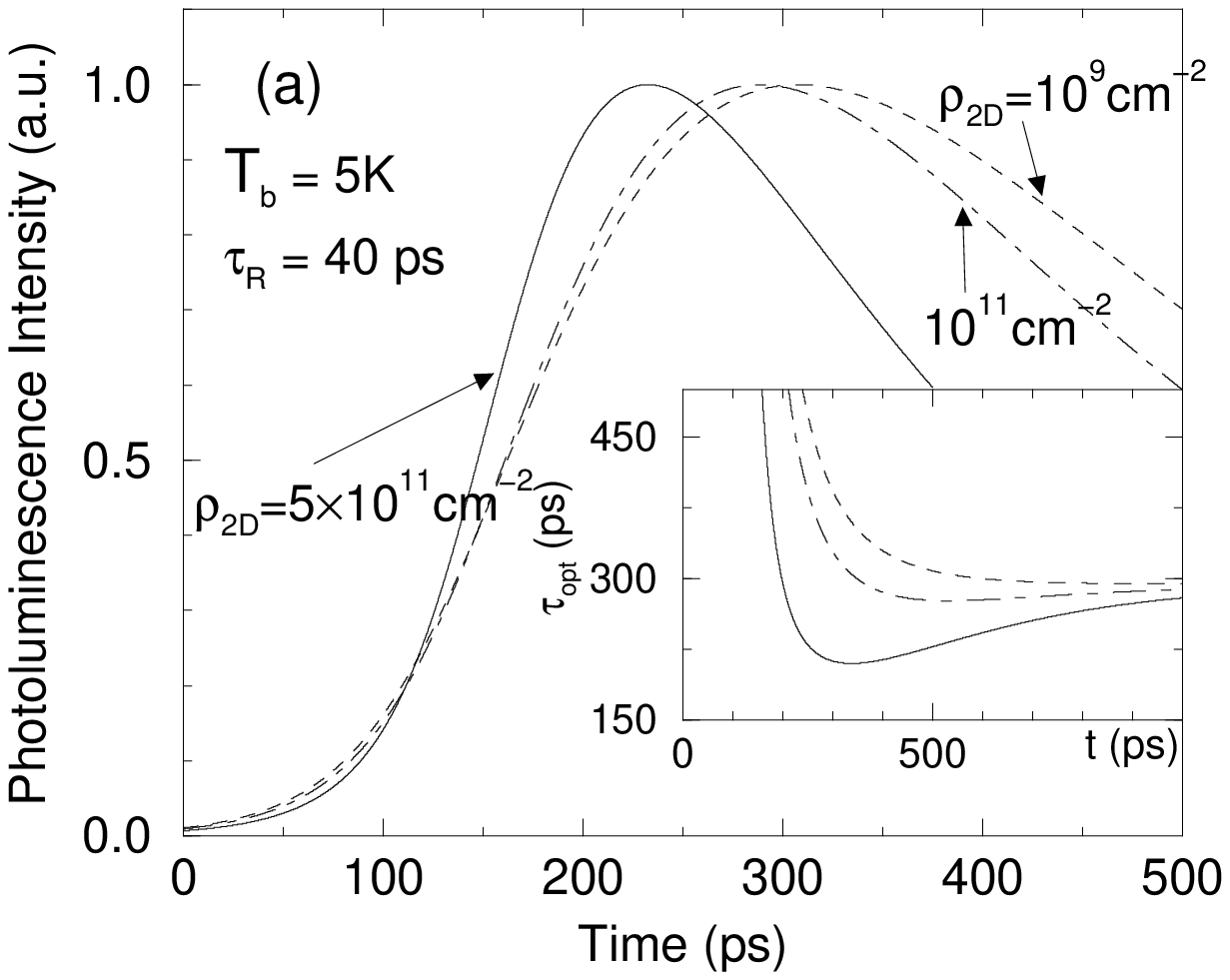}
\end{center}
\begin{center}
\leavevmode
\epsfxsize=8cm \epsfbox{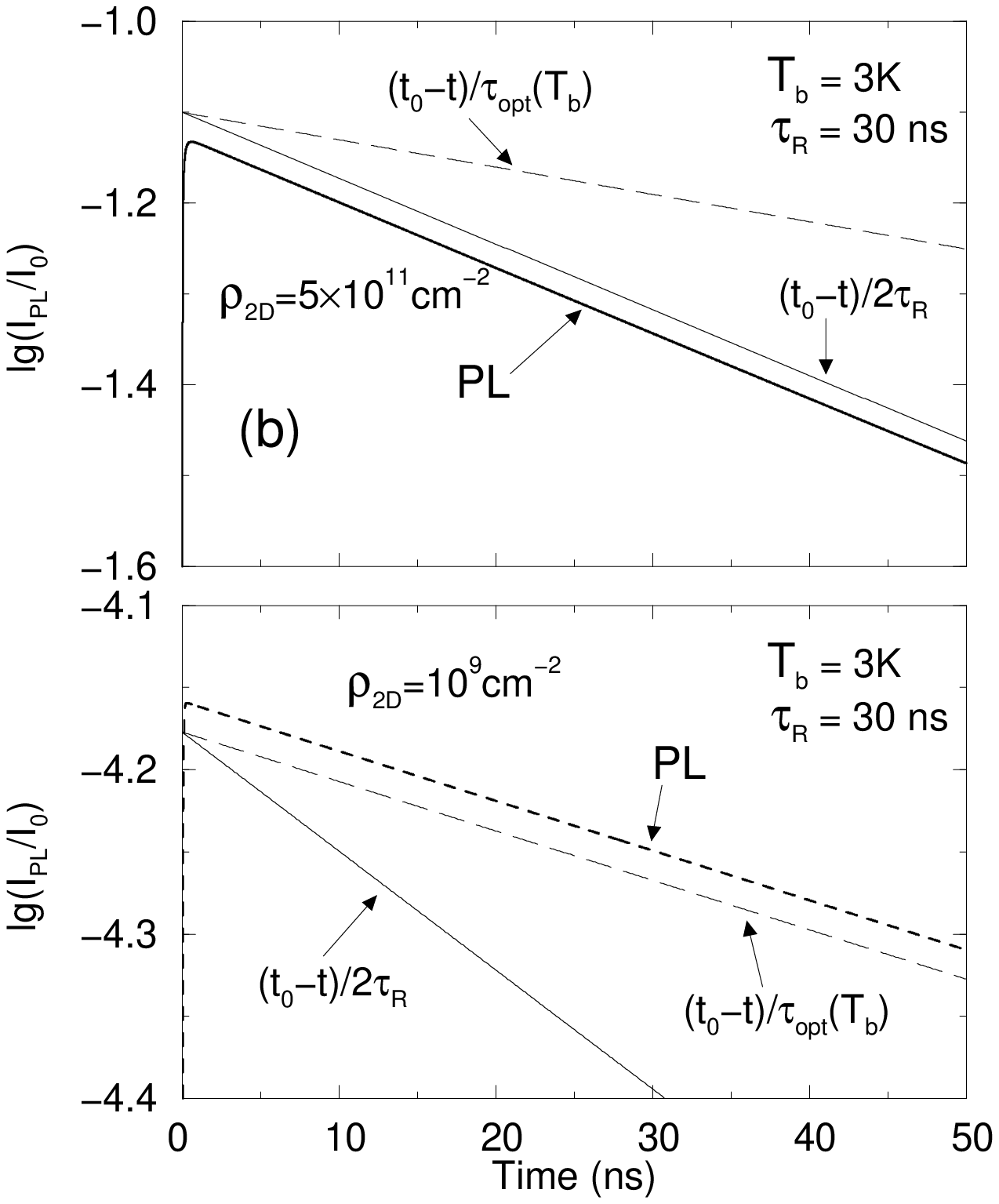}
\end{center}
\caption{ Phototuminescence dynamics of QW excitons modeled with 
Eqs.~(\ref{term6}), (\ref{pl3}), and (\ref{pl6aa})-(\ref{pl6}). (a) GaAs 
QW with $L_z = 100~\AA$, $D = 6.3~eV$, $\tau_{sc} = 85~ns$, $\tau_R = 
40~ps$, $T_b = 5~K$, $\rho_{2D} = 10^9 \ cm^{-2}$ (dashed line), $10^{11} 
\ cm^{-2}$ (dot-dashed line), and $5 \times 10^{11} \ cm^{-2}$ 
(solid line). Inset: change of the optical decay time $\tau_{opt} = 
\tau_{opt}(t)$. (b) GaAs/AlGaAs CQWs with $\tau_{sc} = 13.9~ns$, 
$\tau_R = 30~ns$, $T_b = 3~K$, $\rho_{2D} = 10^9 \ cm^{-2}$ (bold dashed 
line), and $5 \times 10^{11} \ cm^{-2}$ (bold solid line). The long 
dashed and solid straight lines refer to the exponential kinetics given 
by $I_{PL}(t) \propto \exp[-t/\tau_{opt}(T_b)]$ and $I_{PL}(t) \propto 
\exp[-t/(2\tau_R)]$, respectively. The PL intensities $I_{PL}(t)$ of (a) 
and (b) are normalized by $I_0 = 248.31~kW/cm^2$. }
\label{fig9b}
\end{figure}
\section{Discussion}
There are three neglected factors, the interface polariton effect, 
the low-energy biexciton states, and residual interface disorder, 
which can affect the developed model. 

In perfect QWs, the radiative modes $p_{\|} < k(\omega)$, which are 
responsible for the optical decay of QW excitons in bulk photons, are 
accompanied by confined modes $p_{\|} > k(\omega)$ located outside 
the photon cone. These modes give rise to QW interface polaritons. 
While the QW polaritons cannot be seen in standard optical experiments, 
which deal with bulk photons, they can modify the quadratic in-plane 
dispersion of QW excitons and contribute to the total optics in a 
``hidden'' way. Therefore in the general case both conjugated phenomena, 
the radiative decay into bulk photons and the QW polariton effect, 
should be treated simultaneously \cite{Ivanov1}. The dispersion law of 
$T$-polarized QW polaritons is $c^2p_{\|}^2 / \epsilon_b 
= \omega^2 \Big[ 1 + (R_c \sqrt{p_{\|}^2 - \epsilon_b 
\omega^2 / c^2}) / (\omega_t^2 + \hbar \omega_t p_{\|}^2/M_x - i 
\Gamma^x \omega - \omega^2) \Big]$, where $R_c = (\pi e^2 f_{xy}) / 
(\epsilon_b m_0)$, $f_{xy}$ is the 
oscillator strength of exciton-photon coupling per unit area of a QW 
(the subscript $xy$ refers to the in-plane polarization of the light), 
and $\Gamma^x$ is the rate of incoherent scattering of QW excitons 
\cite{Andreani2,Nakayama}. Actually the dispersion equation can be 
applied to the both radiative and confined modes \cite{Ivanov1}. Within 
the photon cone the solution $\omega = \omega(p_{\|})$ is $\omega = 
\omega_t - i\Gamma_T(p_{\|})/2$, where $\Gamma_T(p_{\|})$ is given by 
Eq.~(\ref{pl1}) with $\Gamma_0 = (\sqrt{\epsilon_b}/c) R_c$. The 
dispersion $\omega = \omega_T(p_{\|})$ of $T$-polarized QW polaritons 
is located outside the photon cone, approaches the quadratic 
dispersion $\hbar p_{\|}^2/2M_x$ at $p_{\|} \gg k_0$, and continuously 
transforms to the photon dispersion $c p_{\|}/\sqrt{\epsilon_b}$ at 
$p_{\|} \leq k_0$. Thus during thermalization the QW excitons can 
accumulate at the ``bottleneck'' band of the $T$-polariton dispersion 
rather than in the radiative zone given by $\omega = \omega_t + \hbar 
p_{\|}^2/2M_x$ for $p_{\|} \leq k_0$. 

The polariton effect will not influence the thermalization and PL 
kinetics provided that $E_{k_0} \gg (1/2) \hbar \Omega^{QW}_c$, where 
the effective QW polariton parameter $\Omega^{QW}_c$ is given by 
$(\Omega^{QW}_c)^2 = (\sqrt{\epsilon_b}/c)R_c \omega_t$. In this 
case the QW polariton bottleneck band is very narrow, located at 
energies above $\hbar \omega_t$, and the mode ${\bf p_{\|}}=0$ preserves 
the status of a ground-state mode. For a GaAs QW with $L_z = 100 \ \AA$ 
one has $f_{xy} \simeq 5 \times 10^{-4} \ \AA^{-2}$ \cite{Andreani1}. 
This value of the oscillator strength yields $\hbar^2 R_c \simeq 
1.2 \times 10^{-2} eV^2 \AA$, $\tau_R = 1/\Gamma_0 \simeq 25 \ ps$, 
and $(1/2) \hbar \Omega^{QW}_c \simeq 3 \ meV$, i.e., $(1/2) \hbar 
\Omega^{QW}_c > E_{k_0} \simeq 0.1 \ meV$. In perfect CQWs the oscillator 
strength $f_{xy}$ can be made two or three orders smaller than the used 
above value. This means that indeed $E_{k_0} \gg (1/2) \hbar 
\Omega^{QW}_c$ and one can apply Eqs.~(\ref{term6}), (\ref{pl3}), and 
(\ref{pl6aa})-(\ref{pl6}) to the relaxational and PL dynamics of indirect 
CQW excitons. 

The exciton-exciton interaction relaxes the 
polariton effect in perfect QWs and removes it, if $\Gamma^x = 
1/\tau_{x-x} \geq \Omega^{QW}_c$. With the estimate of $1/\tau_{x-x}$ 
given by Eq.~(A4) of Appendix we conclude that for direct excitons in 
a single GaAs QW with $L_z = 100 \ \AA$ the interface polariton effect 
plays no role at concentrations $\rho_{2D} > 3 \times 10^{10} \ cm^{-2}$. 
Alternatively, in the presence of the well-developed polariton picture 
with $E_{k_0} < (1/2) \hbar \Omega^{QW}_c$, the relaxation kinetics 
should be reformulated in terms of ``QW polariton $\pm$ bulk LA-phonon 
$\leftrightarrow$  QW polariton''. Furthermore, in this case 
relaxational thermodynamics becomes invalid due to the absence of 
steady-state quasi-equilibrium in a gas of QW excitons (QW polaritons). 

Next we consider the formation of quasi-2D biexcitons. The potential 
for exciton-exciton interaction is repulsive for indirect excitons in 
CQWs. However, in a single QW two excitons in some spin configurations, 
e.g., $\sigma^+$- and $\sigma^-$ - spin-polarized, attract each other 
due to the exchange Coulomb interaction. This interaction may lead to 
biexciton states which lie below the excitonic energies $E \geq 0$. 
While the biexciton states are absent in CQWs, they are presented in 
single GaAs QWs at low temperatures and moderate optical excitations. 
Typical values of the biexciton binding energy $E_m^{(2D)}$ are about 
$1 - 2 \ meV$ for $L_z = 200 - 100 \ \AA$. The law of mass action 
\cite{Wolfe} applied to QW excitons and biexcitons, which are 
Maxwell-Boltzmann distributed at the same effective temperature $T$, 
shows that the biexciton states are ionized in unbound excitons at 
$k_BT \geq E_m^{(2D)}$, i.e., at $T \geq 10~K$. At low temperatures, 
with increasing dimensionless parameter $[a_m^{(2D)}]^2 \rho_{2D}$ the 
biexciton binding energy decreases due to screening by 
quasi-equilibrium QW excitons, which remain well-defined 
quasi-particles for $[a_x^{(2D)}]^2 \rho_{2D} \leq 1$. Here $a_m^{(2D)}$ 
($> a_x^{(2D)}$) is the radus of a QW biexciton. The screening of the 
biexciton states is mainly due to the repulsive interaction between QW 
excitons of the identical spin structure, which is much stronger than 
the attractive potential \cite{Ivanov3,Keldysh}. The biexciton state 
is affected by excitons if $[a_m^{(2D)}]^2 \rho_{2D} E_x^{(2D)} \geq 
E_m^{(2D)}$, where $E_x^{(2D)}$ is the two-dimensional excitonic 
Rydberg. This estimate refers to low effective temperatures $T 
\rightarrow 0$. The squared biexciton radius is approximated by 
$[a^{(2D)}_m]^2 = 2 (\mu_x/M_x)(E_x^{(2D)}/ E_m^{(2D)})[a_x^{(2D)}]^2 
\simeq 2.4 [a_x^{(2D)}]^2$, where $\mu_x$ is the reduced exciton mass. 
We estimate that the biexciton state starts to weaken at 
$[a_x^{(2D)}]^2 \rho_{2D} \geq 0.05 - 0.1$, i.e., at $\rho_{2D} \simeq 
10^{11} \ cm^{-2}$. Certainly in the high-density limit $\rho_{2D} 
\simeq 2 - 3 \times 10^{12} \ cm^{-2}$, where the Mott parameter 
$\rho_{2D} [a_x^{(2D)}]^2$ approaches unity, the thermalization and 
photoluminescence of strongly degenerate QW excitons occurs in the 
absence of the biexciton states and can be analyzed with our approach. 
During the optical decay, at concentrations $\rho_{2D} \sim 10^{11} \ 
cm^{-2}$, the QW biexcitons can contribute to the relaxation and PL 
processes. For these densities of QW excitons the relaxation 
thermodynamics given by Eq.~(\ref{term6}) can be generalized to 
include the conversion of QW excitons to QW biexcitons and the 
LA-phonon-assisted relaxation within the both excitonic and 
biexcitonic bands. In this case the quasi-equilibrium concentrations 
of QW excitons and biexcitons are connected by the law of mass action 
\cite{Wolfe}. 

The developed thermalization and photoluminescence kinetics refer to a 
perfect QW. With weak disorder, relaxation can be 
``disorder-assisted'', whereby momentum and energy conservation are 
satisfied by LA-phonon emission concomitant with scattering from the 
disorder potential. This will be relevant for relieving the phonon 
bottleneck if there are significant Fourier components of the disorder 
with wavevectors $p_{d} \geq 2 M_x v_s/\hbar = 1.9 \times 
10^5~cm^{-1}$. For the opposite limit of strong interface and alloy 
disorder, the phonon-assisted relaxation kinetics of localized QW 
excitons is modeled in Ref.~\cite{Takagahara} and more recently in 
Ref.~\cite{Zimmermann} where the PL process is also studied. The 
optical decay of localized QW excitons is investigated in 
Ref.~\cite{Citrin}. Note that the number of localized QW excitons 
saturates due to Pauli blocking, if the concentration of the residual 
defect centers $N_D$ is much less than $\rho_{2D}$ (in perfect GaAs 
QWs $N_D \leq 10^{10} \ cm^{-2}$ \cite{Citrin}). In this case the 
localization processes do not affect the BE kinetics of free QW 
excitons. 

At low bath temperatures $T_b \leq E_0$ the large thermalization times 
$\tau_L \propto \exp(-E_0/k_BT_b)$ and $\tau_{0L} = (T_0/T_b) \tau_L$ 
given by Eqs.~(\ref{exp5}) and (\ref{exp9}) for statistically 
nondegenerate and strongly degenerate QW excitons, respectively, can be 
changed by two-LA-phonon-assisted relaxation processes ``QW exciton $\pm$ 
bulk LA-phonon $\pm$ bulk LA-phonon $\leftrightarrow$ QW exciton''. 
While these processes are of the next order of smallness with respect to 
the considered QW-exciton -- one-phonon scattering, they can be dominant 
in the presence of drastic slowing down of one-phonon thermalization. 
The energy-momentum conservation law in two-phonon-assisted relaxation 
allows for a QW exciton with energy $E \leq E_0$ to be scattered into 
the ground-state mode ${\bf p_{\|}} = 0$ only by at first absorbing of a 
LA-phonon and then re-emitting of another LA-phonon. The rate of the 
anti-Stokes -- Stokes two-phonon scattering into the ground-state 
mode is given by 
\begin{eqnarray}
{ 1 \over \tau^{(2)}_{th} } & = & \Bigg( {2 \pi \over \hbar} \Bigg) 
\sum_{\bf q,q'} \ \Big| {\widetilde M}_{x-ph}({\bf q}) \Big|^2 
\Big| {\widetilde M}_{x-ph}({\bf q'}) \Big|^2 
\nonumber \\ 
& \times &
{ N_{\bf p_{\|}} 
n^{ph}_{\bf q} (n^{ph}_{\bf q'} + 1) \over ( E_{\bf p_{\|}} + 
\hbar \omega^{ph}_{\bf q} - E_{\bf p_{\|} + q} )^2 }
\nonumber \\
& \times &
\delta( E_{\bf p_{\|}} + \hbar \omega^{ph}_{\bf q} - 
\hbar \omega^{ph}_{\bf p_{\|} + q}) \delta_{\bf p_{\|} - q' + q} \ , 
\eqnum{40}
\label{d1} 
\end{eqnarray}
where $E_{\bf p_{\|}} = \hbar^2 p_{\|}^2/2M_x$ and $\hbar 
\omega^{ph}_{\bf q} = \hbar v_s q$ are the energies of a QW exciton 
and a bulk LA-phonon, respectively. The matrix element of QW exciton 
-- bulk LA-phonon interaction is ${\widetilde M}_{x-ph}({\bf q}) = 
[(\hbar D^2 q)/(2v_s \rho V)]^{1/2} F_z(q_zL_z/2)$, where $V$ is the 
volume which refers to bulk LA-phonons. For a quantum gas of excitons 
Eq.~(\ref{d1}) yields the following estimate of the two-phonon-assisted 
relaxation rate: 
\begin{equation}
{ 1 \over \tau^{(2)}_{0L} } = 8 \pi \Bigg( { k_B T_b \over E_0 } \Bigg)^4
{\hbar \over E_0 \tau_{sc}^2} \ .
\eqnum{41}
\label{d2} 
\end{equation}

This result can be motivated as follows. 
Due to the anti-Stokes component, only a small phase space volume given 
by $p_{\|} \leq p^{max}_{\|} = (2k_BT_b)/(\hbar v_s)$ contributes to the 
two-phonon relaxation. The dependence $1 / \tau^{(2)}_{0L} \propto 
T_b^4$ stems from $N_{\bf p_{\|}} n^{ph}_{\bf q} (n^{ph}_{\bf p_{\|}+q} 
+ 1) \propto T_b^3$ and from $p^{max}_{\|} \propto T_b$ on the r.h.s. 
of Eq.~(\ref{d1}). The two-phonon-assisted thermalization time 
$\tau^{(2)}_{0L}$ is proportional to $\tau_{sc}^2$ indicating the next 
order of smallness with respect to $\tau_{0L} \propto \tau_{sc}$. 
Equations (\ref{exp9}) and (\ref{d2}) show that $\tau^{(2)}_{0L}(T_b)$ 
indeed increases with decreasing $T_b$ slowly than $\tau_{0L}(T_b)$, 
i.e., the two-phonon scattering processes become dominant at $0 < T_b 
\leq T_b^c$, where $T_b^c$ is given by the equation $\tau_{0L}(T_b) = 
\tau^{(2)}_{0L}(T_b)$. For example, for $T_b/T_0 = 0.2$ the crossover 
occurs at $T_b^c \simeq 0.06 E_0/k_B = 31.3~mK$. Because 
many-phonon-assisted relaxation of QW excitons at the effective 
temperature $T \leq E_0/k_B$ into the ground-state mode occurs only 
through intermediate anti-Stokes scattering, the relaxation dynamics at 
$T_b = 0$ described by Eq.~(\ref{ground1}) is universal. 

Thus our model refers to indirect excitons in perfect GaAs/AlGaAS CQWs 
and to direct excitons in single QWs. In the latter case, however, the 
concentration of QW excitons should be $\rho_{2D} \geq 10^{11} \ 
cm^{-2}$ in order to suppress the interface polariton and biexciton 
effects. The equations~(\ref{term6}), (\ref{pl3}), and 
(\ref{pl6aa})-(\ref{pl6}) provide us with an universal description 
of the thermalization and photoluminescence kinetics of QW excitons 
within the thermodynamic picture. Note, that the concrete 
characteristics of QW exciton -- bulk LA-phonon interaction enter the 
basic thermodynamic Eq.~(\ref{term6}) only through the elementary 
scattering time $\tau_{sc}$ and the form-factor function $F_z(\chi)$. 
Our approach can also be applied to QW magneto-excitons. In this case 
the quantum-statistical effects will be considerably enhanced, because 
the spin degeneracy factor $g$ of Eq.~(\ref{intr1}) reduces from $g=4$ 
to $g=1$. 
\section{Conclusions}
In this paper we have studied the influence of Bose-Einstein  
statistics on the thermalization, radiative decay, and 
photoluminescence kinetics of a gas of QW excitons. The numerical 
calculations presented in the work deal with perfect GaAs/AlGaAs 
quantum wells. The following conclusions summarize our results. 

(i) We have developed relaxational thermodynamics for QW excitons 
coupled to bulk thermal LA-phonons. The thermodynamic picture implies 
that the concentration $\rho_{2D}$ of QW excitons is larger than the 
critical density $\rho_{2D}^c$ (for GaAs QWs with $L_z = 100~\AA$ we 
estimate $\rho_{2D}^c \simeq 1.2 \times 10^9 \ cm^{-1}$). The 
relaxational thermodynamics is given by Eq.~(\ref{term6}) for the 
effective temperature $T(t)$ of QW excitons. This equation describes 
the thermalization kinetics of a quasi-equilibrium gas of QW excitons 
from the initial temperature $T_i$ to the bath $T_b$. For a quantum 
degenerate gas of quasi-2D excitons, when $T$ is less than the 
degeneracy temperature $T_0$, BE statistics yields nonexponential and 
density-dependent thermalization. In particular, for low bath 
temperatures $T_b \leq E_0/k_B \sim 1 \ K$ the relaxation processes in 
a degenerate gas of QW excitons slow down dramatically and the 
thermalization law is given by $T(t) \propto 1/\ln t$. 

(ii) The effective radiative lifetime $\tau_{opt} = \tau_{opt}(T,T_0)$ 
of quasi-equilibrium BE-distributed QW excitons is given by 
Eq.~(\ref{pl3}). The temperature-independent corrections, given by 
Eq.~(\ref{pl6a}), to the well-known classical limit $\tau_{opt} \propto 
T$ show that BE statistics can be traced in the optical decay of a gas 
of QW excitons even at high effective temperatures $T$. Nonclassical 
statistics strongly influences the optical decay at $T \leq T_0$. In 
particular, at $T \rightarrow 0$ the lifetime $\tau_{opt}$ approaches 
$2 \tau_R$ rather than zero as in the classical limit $\tau_{opt} 
\propto T$. Here the intrinsic radiative lifetime $\tau_R$ is 
determined by the oscillator strength $f_{xy}$ of exciton - photon 
coupling in a QW. 

(iii) The PL kinetics of QW excitons is modeled by three coupled 
Eqs.~(\ref{term6}), (\ref{pl3}), and (\ref{pl6aa})-(\ref{pl6}) for 
the effective temperature $T(t)$, effective radiative time 
$\tau_{opt}(T,\rho_{2D}) = 1/\Gamma_{opt}(T,\rho_{2D})$, and 
concentration $\rho_{2D}(t)$, respectively. This approach describes 
within the thermodynamic picture both the rise of PL, due to 
thermalization, and the decay of PL, due to the radiative recombination 
of thermalized QW excitons. For a classical gas of QW excitons 
the PL kinetics depends on the effective temperature $T$, but 
is independent of $\rho_{2D}$. With increasing density $\rho_{2D}$ of 
QW excitons the quantum-statistical effects builds up and the PL 
kinetics becomes density-dependent. 
\acknowledgements 
We appreciate valuable discussions with P. Stenius, L.V. Butov, and 
H. Kalt. Support of this work by the EPSRC (UK) and by the 
DFG-Schwerpunktprogramm ``Quantenkoh\"arenz in Halbleitern'' 
(Germany) is gratefully acknowledged. 
\appendix
\section{Equilibration dynamics of QW excitons}
The equilibration kinetics of QW excitons is mainly due to 
particle-particle scattering and refers to the Boltzmann-Uhlenbeck 
equation 
\begin{eqnarray} 
{\partial \over \partial t} N_{\bf k_{\|}} & = & - { 2 \pi \over 
\hbar } \ \left({U_0 \over S} \right)^2
\sum_{\bf p_{\|},q_{\|}} \Big[ N_{\bf k_{\|}} 
N_{\bf p_{\|}} (1 + N_{\bf p_{\|}-q_{\|}}) 
\nonumber \\ 
&&  \! \! \! \! \! \! \! \! \! \! \! \! \! \! \! \! \! \! \times  
(1 + N_{\bf k_{\|}+q_{\|}})
- (N_{\bf k_{\|}} + 1)(N_{\bf p_{\|}} + 1) 
N_{\bf p_{\|}-q_{\|}} 
N_{\bf k_{\|}+q_{\|}} \Big] 
\nonumber \\
& \times &  
\delta(E_{\bf p_{\|}-q_{\|}} + 
E_{\bf k_{\|}+q_{\|}} - E_{\bf k_{\|}} - E_{\bf p_{\|}}) \ , 
\eqnum{A1}
\end{eqnarray}
where $U_0/S$ is the potential of exciton-exciton Coulombic 
interaction in a QW, $S$ is the area. 
For CQWs this potential is repulsive and 
independent of the spin structure of interacting excitons owing  
to a rather small contribution of the exchange interaction. For 
a single QW the potential depends upon the spin state of 
excitons and can be repulsive (e.g., for the QW excitons with an 
identical spin structure) as well as attractive (e.g., for 
$\sigma^+$- and $\sigma^-$- spin polarized excitons) \cite{Ivanov3}. 
However, the repulsive potential strongly dominates \cite{Keldysh} and 
determines the equilibration dynamics of QW excitons at $\rho_{2D} 
\geq \rho^c_{2D}$. 

In order to evaluate a characteristic equilibration time $\tau_{x-x}$ 
we analyze within Eq.~(A1) the dynamics of a small fluctuation 
$\delta N_0^{(0)} = \delta N_0(t=0) \ll N_{{\bf p_{\|}}=0}^{eq}$ 
of the ground-state mode population, provided that the occupation 
numbers of all other modes are given by the BE distribution of 
Eq.~(\ref{term3}). The linearization of the r.h.s. of Eq.~(A1) with 
respect to $\delta N_0$ yields an exponential law for the decay of the 
fluctuation, i.e., $d (\delta N_0) / dt = - \delta N_0 / \tau_{x-x}$ 
and $\delta N_0(t) = \delta N_0^{(0)} \exp(-t/\tau_{x-x})$. 
The decay time of the fluctuation, which we attribute to the 
characteristic equilibration time, is given by 
\begin{eqnarray} 
{1 \over \tau_{x-x}} & = & \left( {U_0 M_x \over 2 \pi} \right)^2 \
\left( { k_B T \over \hbar^5 } \right) \ e^{-T_0/T} \ \left( 1 - 
e^{-T_0/T} \right) 
\nonumber \\
& \times &
\int_0^{\infty} du \int_0^{2 \pi} d \phi 
{ e^{2u} \over \Big[ e^{u(1 - \cos \phi)} + e^{-T_0/T} - 1 
\Big] } 
\nonumber \\ 
& \times & 
{1 \over  \Big[ e^{u(1 + \cos \phi)} + e^{-T_0/T} - 1 \Big]
\Big[ e^{2u} + e^{-T_0/T} - 1 \Big] } \ . 
\nonumber \\
\eqnum{A2}
\end{eqnarray}
In order to derive Eq.~(A2) from Eq.~(A1) we put ${\bf k_{\|}} = 0$ 
and introduce new dummy variable ${\bf q_{\|}}' = {\bf q_{\|}} - 
{\bf p_{\|}}/2$. The dimensionless integration variable $u$ on the 
r.h.s. of Eq.~(A2) is given by $u = (\hbar p_{\|})^2/(4 k_B M_x T)$, 
$\phi$ is the angle between ${\bf q_{\|}}'$ and ${\bf p_{\|}}$.

While the repulsive potential $U_0$ for QW excitons in an identical 
spin state can be explicitly written in terms of an integral convolution 
of the different pair Coulomb potentials between the constituent 
electrons and holes with the excitonic wavefunctions (see, e.g., 
Ref.~\cite{SchmittRink}), here we give a scaling estimate of 
$U_0$. The potential $U_0$ is determined by $U_0 = \int U_{x-x}(r_{\|}) 
d{\bf r_{\|}}$, where $U_{x-x}(r_{\|})$ is the real-space potential 
of exciton-exciton interaction and $r_{\|}$ is the distance between 
two interacting QW excitons. The characteristic energy and length scales 
for $U_{x-x}(r_{\|})$ are given by the two-dimensional Rydberg 
$E_x^{(2D)} = 2 \mu_x e^4 / \hbar^2$ and the corresponding excitonic Bohr 
radius $a_x^{(2D)} = \hbar^2 / (2 \mu_x e^2)$, where $\mu_x$ is the 
reduced mass of a QW exciton. For simple model potentials 
$U_{x-x}(r_{\|}) = C E_x^{(2D)} \exp(-r_{\|}/a_x^{(2D)})$ and 
$U_{x-x}(r_{\|}) = C E_x^{(2D)} (a_x^{(2D)} / r_{\|}) 
\exp(-r_{\|}/a_x^{(2D)})$, where the constant $C \simeq 1$, one estimates 
\begin{equation}
U_0 = 2 \pi C \ E_x^{(2D)} [a_x^{(2D)}]^2 = \pi C \ {\hbar^2 \over 
\mu_x} \ . 
\eqnum{A3}
\end{equation}
A remarkable feature of Eq.~(A3) is that while with decreasing confinement 
in the $z$-direction (the QW growth direction) $E_x^{(2D)}$ and 
$a_x^{(2D)}$ start to approach the corresponding bulk values, their 
combination $E_x^{(2D)} [a_x^{(2D)}]^2$ does not change. Such behavior 
of $U_0$ is due to the quasi-two-dimensionality of QW excitons. The 
constant $C$, which is of the order of unity, depends on the design of a 
QW, the shape of the model potential $U_{x-x}(r_{\|})$, etc.. 
In further analysis we put $C=1$. 

For a classical gas of QW excitons, when $T \gg T_0$, one obtains 
from Eqs.~(A2)-(A3): 
\begin{equation}
{1 \over \tau_{x-x}} = { \pi \over 4 \hbar } 
\left( { M_x \over \mu_x } \right)^2 k_B T_0 \ ,  
\eqnum{A4}
\end{equation}
where $T_0$ is given by Eq.~(\ref{intr1}). According to Eq.~(A5), the 
scattering rate due to particle-particle interaction is proportional 
to the concentration $\rho_{2D}$ of QW excitons. Furthermore, as a 
signature of the quasi-two-dimensionality of excitons in QWs, the 
characteristic equilibration time $\tau_{x-x}$ is independent of the 
temperature $T$ (for $T \gg T_0$) and of the scattering length $\sim 
a_x^{(2D)}$. 

For a quantum gas of QW excitons ($T \ll T_0$) we find from 
Eqs.~(A2)-(A3): 
\begin{equation}
{1 \over \tau_{x-x}} = 
{ \pi \over \hbar } \left( 1 - {\pi \over 4} \right) 
\left( { M_x \over \mu_x } \right)^2 k_B T \ e^{T_0/T} \ . 
\eqnum{A5}
\end{equation}
The considerable increase of the scattering rate given by Eq.~(A5) 
in comparison with that of Eq.~(A4) is due to the low-energy QW excitons 
with $N^{eq}_E \gg 1$ at $E \leq k_B T$. 
\section{Homogeneous broadening in the radiative decay of QW excitons}
The joint density of states $J(p_{\|})$ for the resonant optical decay 
of a QW exciton with $p_{\|} \leq k_0$ into a bulk photon is given by  
\begin{eqnarray}
J(p_{\|}) & = & \Bigg( {1 \over 2 \pi^2} \Bigg) 
\int^{+\infty}_{-\infty} dk_{\perp} 
\nonumber \\
& \times & 
{ \Gamma_{hom} \over \Gamma_{hom}^2 + \Big[ \omega_x(p_{\|}) - 
\omega_{\gamma}(p_{\|},k_{\perp}) \Big]^2 } \ ,
\eqnum{B1}
\end{eqnarray}
where $\hbar \omega_x(p_{\|}) = \hbar \omega_t + \hbar^2 p_{\|}^2 / 
2 M_x$ and $\omega_{\gamma}(p_{\|},k_{\perp}) = 
(c/\sqrt{\epsilon_b}) \sqrt{p_{\|}^2 + k_{\perp}^2}$ are the 
dispersions of QW excitons and bulk photons, respectively, $k_{\perp}$ 
is the $z$-component of the wavevector of bulk photons, which  
resonantly interact with a QW exciton with in-plane wavevector $p_{\|}$, 
and $\Gamma_{hom}$ is the homogeneous linewidth of QW excitons. 
If $\partial \omega_{\gamma}(p_{\|},k_{\perp}) / \partial k_{\perp} 
\neq 0$ at $k_{\perp}$ determined by the equation $\omega_x(p_{\|}) = 
\omega_{\gamma}(p_{\|},k_{\perp})$, the integrand on the r.h.s. 
of Eq.~(B1) preserves its Lorentzian shape in wavevector coordinates 
and the joint density of states $J(p_{\|})$ is independent of 
$\Gamma_{hom}$. This is a regular case, which does not hold for $p_{\|} 
\rightarrow k_0$ ($p_{\|} \leq k_0$). In this case the solution of 
$\omega_x(k_0) = \omega_{\gamma}(k_0,k_{\perp})$ is $k_{\perp} = 0$ 
and $\partial \omega_{\gamma}(p_{\|}=k_0,k_{\perp}) / \partial k_{\perp} 
= 0$, indicating 1D van Hove singularity in the joint density of 
states. As a result, $J(p_{\|})$ becomes $\Gamma_{hom}$-dependent 
in a close vicinity of $p_{\|} = k_0$. This is the only way that the 
homogeneous linewidth influences the optical decay of quasi-equilibrium 
excitons in ideal QWs. 

With substitution $k_{\perp} = p_{\|} \tan \phi$, where $\phi \ni 
[-\pi/2,\pi/2]$, Eq.~(B1) reduces to 
\begin{eqnarray}
J(p_{\|}) & = & {1 \over \pi^2} \Bigg[{p_{\|} \over \omega_x(p_{\|})} 
\Bigg] \int^{\pi/2}_{0} du 
\nonumber \\ 
& \times & { {\tilde \gamma} \over (1/4)u^4 - 
({\tilde \delta}^2/2)u^2 + ({\tilde \delta}^4/4 + \delta^2 {\tilde 
\gamma}^2) } \ , 
\eqnum{B2}
\end{eqnarray}
where ${\tilde \gamma} = \Gamma_{hom}/\omega_t$, $\delta = 
\cos \phi_0 = p_{\|}/k_0 = (c p_{\|})/ (\epsilon_b \omega_t)$, 
${\tilde \delta} = \sin \phi_0 = \sqrt{1 - \delta^2}$, and 
the integration variable $u = \phi - \phi_0 + \sin \phi_0$. 
Straightforward integration of Eq.~(B2) yields the joint density 
of states: 
\begin{eqnarray}
J(p_{\|}) & = & {\sqrt{2} \over \pi} \ \Bigg[{p_{\|} \over 
\omega_x(p_{\|})} \Bigg] \ { {\tilde \gamma} \over 
\Big( 4 \delta^2 {\tilde \gamma}^2 + {\tilde \delta}^4 \Big)^{1/2} } 
\nonumber \\ 
& \times & 
{ 1 \over \Big[ \Big( 4 \delta^2 {\tilde \gamma}^2 + 
{\tilde \delta}^4 \Big)^{1/2} - {\tilde \delta}^2 \Big]^{1/2} } \ . 
\eqnum{B3}
\end{eqnarray}
For ${\tilde \delta} \gg \sqrt{2 {\tilde \gamma}}$, which is identical 
to the condition $k_0 - p_{\|} \gg {\tilde \gamma} k_0$, 
Eq.~(B3) reduces to the standard formula \cite{Andreani1}:
\begin{eqnarray}
J(p_{\|} < k_0 - {\tilde \gamma} k_0) & = & {1 \over \pi} 
\Bigg( {\epsilon_b \over c^2} \Bigg) { \omega_t \over 
\sqrt{k_0^2 - p_{\|}^2} } \ .   
\eqnum{B4}
\end{eqnarray}
In opposite limit ${\tilde \delta} \ll \sqrt{2 {\tilde \gamma}}$, 
i.e., for $p_{\|} \rightarrow k_0$, Eq.~(B3) yields 
\begin{equation}
J(p_{\|}=k_0) = { 1 \over 2 \pi} \  
{ k_0 \over (\Gamma_{hom} \omega_t)^{1/2} } \ .
\eqnum{B5}
\end{equation}
Equation (B5) shows how the homogeneous linewidth $\Gamma_{hom}$ 
removes the van Hove singularity at $p_{\|}=k_0$. For ${\tilde \delta} 
\leq \sqrt{2 {\tilde \gamma}}$ the joint density of states is 
$\Gamma_{hom}$-dependent. 

With Eq.~(B3) we derive the final expression for the intrinsic 
radiative rate of $T$-polarized QW excitons in the presence of 
homogeneous broadening: 
\begin{eqnarray}
&& \Gamma_T(p_{\|}) = { \Big( \sqrt{2} k_0^2 p_{\|} {\tilde \gamma} 
\Big) \ \Gamma_0 \over \Big[ (k_0^2 - p_{\|}^2)^2 + 4 {\tilde 
\gamma}^2 k_0^2 p_{\|}^2 \Big]^{1/2} } 
\nonumber \\ 
&& \times {1 \over 
\Bigg[ \Big[ (k_0^2 - p_{\|}^2)^2 
+ 4 {\tilde \gamma}^2 k_0^2 p_{\|}^2 \Big]^{1/2} - (k_0^2 - p_{\|}^2) 
\Bigg]^{1/2} } \ . 
\eqnum{B6}
\end{eqnarray}
For $k_0 - p_{\|} \gg {\tilde \gamma} k_0$ 
Eq.~(B4) is identical to Eq.~(\ref{pl1}). The homogeneous linewidth 
$\Gamma_{hom}$ changes Eq.~(\ref{pl1}) for $\Gamma_T(p_{\|})$ only 
at the band of states given by ${\tilde \gamma} k_0 > k_0 - p_{\|} 
\geq 0$ and, in particular, $\Gamma_T(p_{\|}=k_0) = \Gamma_0/(2 
{\tilde \gamma}^{1/2}$) rather than divergent. Note that the band 
of states is rather narrow because ${\tilde \gamma} \simeq 1/(\tau_{x-x} 
\omega_t) \sim 10^{-3}$. The 1D van Hove singularity at $p_{\|} = k_0$ 
is absent for the $L$-polarized QW excitons. 

Homogeneous broadening practically does not change the optical decay 
rate of a gas of QW excitons. The use of Eq.~(B6) for calculation of 
$\Gamma_{opt}$ by Eq.~(\ref{pl2}) shows that for Maxwell-Boltzmann 
distributed QW excitons one obtains only a small renormalization of 
$J_T$ of Eq.~(\ref{pl3}), i.e., $J_T \rightarrow J_T (1 - {\tilde 
\gamma}^{1/2})$. As a result, the optical decay rate of a classical gas 
of QW excitons changes from $\Gamma_{opt}$ to $\Gamma_{opt}[1 - 
(3/4){\tilde \gamma}^{1/2}]$. This correction due to the homogeneous 
linewidth is very small and can indeed be neglected. The corresponding 
correction for statistically degenerate QW excitons is even smaller 
because the low-energy excitons tend to populate the ground-state mode 
${\bf p}_{\|} = 0$ rather than vicinity of the energy states 
$E \simeq E_{k_0}$. 

\begin{references}
\bibitem{Masumoto} Y. Masumoto, S. Shionoya, and H. Kawaguchi, 
Phys. Rev. B {\bf 29}, 2324 (1984). 
\bibitem{Koteles} J. Lee, E.S. Koteles, and M.O. Vassell, 
Phys. Rev. B {\bf 33}, 5512 (1986). 
\bibitem{Feldmann} J. Feldmann, G. Peter, E.O. G\"obel, P. Dawson, 
K. Moore, C. Foxon, and R.J. Elliott, Phys. Rev. Lett. {\bf 59}, 
2337 (1987).
\bibitem{Damen} T.C. Damen, J. Shah, D.Y. Oberli, D.S. Chemla, 
J.E. Cunningham, and J.M. Kuo, Phys. Rev. B {\bf 42}, 7434 (1990); 
P.W.M. Blom, P.J. van~Hall, C. Smit, J.P. Cuypers, and J.H. Wolter, 
Phys. Rev. Lett. {\bf 71}, 3878 (1993). 
\bibitem{Deveaud1} B. Deveaud, F. Cl\'erot, N. Roy, K. Satzke, 
B. Sermage, and D.S. Katzer, Phys. Rev. Lett. {\bf 67}, 2355 
(1991). 
\bibitem{Vinattieri} M. Gurioli, A. Vinattieri, M. Colocci, 
C. Deparis, J. Massies, G. Neu, A. Bossachi, and S. Franchi, 
Phys. Rev. B {\bf 44}, 3115 (1991). 
\bibitem{Kuhl} R. Eccleston, R. Strobel, W.W. R\"uhle, J. Kuhl, 
B.F. Feuerbacher, and K. Ploog, Phys. Rev. B {\bf 44}, 1395 (1991); 
R. Eccleston, B.F. Feuerbacher, J. Kuhl, W.W. R\"uhle, and K.~Ploog, 
Phys. Rev. B {\bf 45}, 11403 (1992). 
 \bibitem{Wood} V. Srinivas, J. Hryniewicz, Y.J. Chen, and 
C.E.C. Wood, Phys. Rev. B {\bf 46}, 10193 (1992). 
\bibitem{Weimann} J. Martinez-Pastor, A. Vinattieri, L. Carraresi, 
M. Colocci, Ph. Roussignol, and G.~Weimann, 
Phys. Rev. B {\bf 47}, 10456 (1993). 
\bibitem{Shah0} A. Vinattieri, J. Shah, T.C. Damen, D.S. Kim, 
L.N. Pfeiffer, M.Z. Maialle, and L.J.~Sham, 
Phys. Rev. B {\bf 50}, 10868 (1994). 
\bibitem{Shah1} J. Shah, {\it Ultrafast Spectroscopy of Semiconductors 
and Semiconductor Nano-structures}, Springer Series in Solid-State 
Sciences, Vol.~115 (Springer, Berlin, 1996). 
\bibitem{Axt} S. Grosse, R. Arnold, G. von Plessen, M. Koch, 
J. Feldmann, V.M. Axt, T. Kuhn, R.~Rettig, and W. Stolz, 
Phys. Status Solidi B {\bf 204}, 147 (1997); 
M. Gulia, F. Rossi, E.~Molinari, P.E. Selbmann, and P. Lugli, 
Phys. Rev. B {\bf 55}, R16049 (1997). 
\bibitem{Butov1} L.V. Butov, A. Imamoglu, A.V. Mintsev, 
K.L. Campman, and A.C. Gossard, preprint (submitted to 
Phys. Rev. Lett., 1998). 
\bibitem{Kalt} M. Umlauff, J. Hoffmann, H. Kalt, W. Langbein, 
J.M. Hvam, M. Scholl, J. S\"ollner, M. Heuken, B. Jobst, and 
D. Hommel, Phys. Rev. B {\bf 57}, 1390 (1998); 
J. Hoffmann, M. Umlauff, H. Kalt, W. Langbein, and J.M. Hvam, 
Phys. Status Solidi B {\bf 204}, 195 (1997). 
\bibitem{Takagahara} T. Takagahara, Phys. Rev. B {\bf 31}, 6552 
(1985). 
\bibitem{Andreani1} L.C. Andreani, F. Tassone, and F. Bassani, Solid 
State Commun. {\bf 77}, 641 (1991).
\bibitem{Per1} W. Zhao, P. Stenius, and A. Imamoglu, Phys. Rev. 
{\bf B56}, 5306 (1997). 
\bibitem{Wolfe} J.C. Kim and J.P. Wolfe, Phys. Rev. B {\bf 57}, 
9861 (1998). 
\bibitem{Andreani2} L.C. Andreani and F. Bassani, Phys. Rev. B 
{\bf 41}, 7536 (1990). 
\bibitem{Per2} P. Stenius and A.L. Ivanov, Solid State Commun. 
{\bf 108}, in print (1998); P. Stenius, PhD thesis, 
University of California, Santa Barbara, 1998.  
\bibitem{Bockelmann} U. Bockelmann, Phys. Rev. B {\bf 50}, 
17271 (1994). 
\bibitem{Levich} E. Levich and V. Yakhot, Phys. Rev. {\bf 15}, 
243 (1977). 
\bibitem{Landau} E.M. Lifshitz and L.P. Pitaevskii, {\it Course
of Theoretical Physics} (Pergamon, Oxford, 1980), Vol. 9, Part II,
Sec. 6 and Sec. 25. 
\bibitem{Ivanov0} A.L. Ivanov, C.~Ell, and H. Haug, 
Phys. Rev. E {\bf 55}, 6363 (1997).
\bibitem{Ivanov2} A.L. Ivanov, C.~Ell, and H. Haug, Phys. Status 
Solidi B {\bf 206}, 235 (1998);  C.~Ell, A.L.~Ivanov, and H. Haug, 
Phys. Rev. B {\bf 57}, 9663 (1998). 
\bibitem{Butov3} L.V. Butov, private communication. 
The preliminary result is that a high-density gas of excitons in 
GaAs/AlGaAs CQWs does not cool below $500 \ mK$ within the lifetime 
of indirect QW excitons. In these experiments, the bath temperature is 
$T_b \simeq 50 \ mK$, the initial effective temperature of excitons 
is $T_i \simeq 1 \ K$, and the lifetime is about $30-100 \ ns$. 
\bibitem{Littlewood1} X. Zhu, P.B. Littlewood, M. Hybertsen, 
and T. Rice, Phys. Rev. Lett. {\bf 74}, 1633~(1995).
\bibitem{Littlewood2} P.B. Littlewood and X.J. Zhu, Physica Scripta 
{\bf T68}, 56 (1996). 
\bibitem{Popov} V.N. Popov, Theor. Math. Phys. {\bf 11}, 565 (1972). 
\bibitem{Butov2} T. Fukuzawa, E.E. Mendez, and J. Hong, 
Phys. Rev. Lett. {\bf 64}, 3066 (1990); L.V.~Butov, A. Zrenner, 
G. Abstreiter, G. B\"ohm, and G. Weimann, Phys. Rev. Lett. {\bf 73}, 
304 (1994); L.V. Butov and A.I. Filin, Phys. Rev. B {\bf 58}, 
1980 (1998). The experimental results of the work by T. Fukuzawa 
{\it et al.} were later re-interpreted, see J.A. Kash, M. Zachau, 
E.E. Mendez, J.M. Hong, and T. Fukuzawa, Phys. Rev. Lett. {\bf 66}, 
2247 (1991). 
\bibitem{Golub} J.E. Golub, K. Kash, J.P. Harbison, and 
L.T. Florez,  Phys. Rev. B {\bf 41}, 8564 (1990); 
A.~Alexandrou, J.A. Kash, E.E. Mendez, M. Zachau, J.M. Hong, 
T. Fukuzawa, and Y.~Hase, Phys. Rev. B {\bf 42}, 9225 (1990). 
\bibitem{Nakayama} M. Nakayama, Solid State Commun. 
{\bf 55}, 1053 (1985). 
\bibitem{Ivanov1}  A.L Ivanov, H. Wang, J. Shah, T.C. Damen, 
L.V. Keldysh, H. Haug, and L.N. Pfeiffer, Phys. Rev. B {\bf 56}, 
3941 (1997). 
\bibitem{Ivanov3} A.L. Ivanov, L.V. Keldysh, and V.V. Panashchenko, 
Zh. Eksp. Teor. Fiz {\bf 99}, 641 (1991) [Sov. Phys. JETP {\bf 72}, 
359 (1991)]. 
\bibitem{Keldysh} L.V. Keldysh, Solid. State Commun. {\bf 84}, 37 
(1992);  Phys. Stat. Solidi B {\bf 173}, 119 (1992). 
\bibitem{Zimmermann} R. Zimmermann and E. Runge, Phys. Status Solidi 
A {\bf 164}, 511 (1997); E. Runge and R. Zimmermann, Phys. Status 
Solidi B {\bf 206}, 167 (1998). 
\bibitem{Citrin} D.S. Citrin, Phys. Rev. B {\bf 47}, 3832 (1993). 
\bibitem{SchmittRink} H. Haug and S. Schmitt-Rink, Progr. Quantum 
Electron. {\bf 9}, 3 (1984). 
%
\end{references}
\end{document}